\documentclass[aps,prx,twocolumn,floatfix,english,showpacs,10pt,
superscriptaddress,longbilbiography]{revtex4-2}
\usepackage{amsmath,bm}
\usepackage{mathtools}
\usepackage{commath}
\usepackage{enumitem} 
\usepackage{amssymb}
\usepackage{graphicx}
\usepackage{babel}
\usepackage{cases}
\usepackage[colorlinks=true, citecolor=blue, anchorcolor=green]{hyperref}
\hypersetup{linkcolor=red}
\usepackage{floatrow}
\usepackage{dblfloatfix}
\usepackage{xcolor}
\usepackage{braket}

\makeatletter

\newcommand{\Rmnum}[1]{\expandafter\@slowromancap\romannumeral #1@}

\newcommand{\sub}{\text{sub}}

\newcommand{\tr}{\operatorname{Tr}}

\newcommand{\rank}{\operatorname{rank}}

\makeatother

\begin{document}
\title{Typical Output States of Monitored Random Clifford Circuits: \\
A Graph-Theoretic Approach}
\author{Yu-Xuan Zhang}
\affiliation{Institute of Physics, Chinese Academy of Sciences, Beijing 100190, China}
\affiliation{School of Physical Sciences, 
University of Chinese Academy of Sciences, Beijing 100049, China}

\author{Yu-Xiang Zhang}
\email{iyxz@iphy.ac.cn}
\affiliation{Institute of Physics, Chinese Academy of Sciences, Beijing 100190, China}
\affiliation{School of Physical Sciences, 
University of Chinese Academy of Sciences, Beijing 100049, China}
\date{\today}

\begin{abstract}
In many-body physics, an explicit wavefunction often provides the most thorough understanding, yet
for monitored random Clifford circuit it has remained missing. Here we 
develop a framework that grants direct access to the
typical output states. 
Our formalism is based on the fact that every stabilizer state is 
local-Clifford-equivalent to
a graph state, whose adjacency matrix, a classical bit matrix,
provides a complete description of the quantum state. 
In the large-$N$ limit, we show that these graphs 
converge to the 
Erd\H{o}s--R\'{e}nyi random graph $G(N,1/2)$, which allows us to resolve the
open problem of Greenberger--Horne--Zeilinger (GHZ) 
entanglement generated by deep random Clifford circuits. We obtain 
analytically the mean GHZ content $\langle g_3\rangle=1.204$ for 
even $N$ and $1.325$ for odd $N$. 
For monitored Clifford circuits with a one-dimensional brickwork layout, 
we uncover an emergent
Erd\H{o}s--R\'{e}nyi subgraph $G(N_{\mathrm{sub}},1/2)$ in the output states
of the volume-law phase,
where $N_{\sub}/N\approx \sqrt{1-p/p_c}$ with $p$ the measurement rate and $p_c$
the critical point of the measurement-induced phase transition (MIPT).
The output state is thus equivalent to the output of an unmonitored 
random Clifford circuit 
on $N_{\mathrm{sub}}$ qubits, weakly
perturbed by the remaining $N-N_{\mathrm{sub}}$ qubits carrying little entanglement.
This result directly accounts for the quantum error-correcting capability of
the volume-law phase, and implies the same GHZ statistics for
the whole volume-law phase. We further identify a 
clustering effect for qubits in the dense subgraph, which we 
reproduce with an infection-recovery toy model
that exhibits a measurement-induced absorbing-state phase transition. 
Finally, a mean-field argument on the graph locates the MIPT critical point at 
$p_c = 0.1608$, in excellent agreement with the numerical value $p_c\approx 0.16$. 
\end{abstract}

\maketitle

\section{Introduction}

Quantum circuits with random local Clifford gates and 
Pauli measurements provide a versatile platform for investigating 
non-equilibrium quantum many-body dynamics using
information-theoretical methods~\cite{Fisher:2023aa}.
They can be simulated efficiently by classical means
due to the Gottesman-Knill theorem~\cite{Gottesman:1998aa,Aaronson:2004aa,Anders:2006aa},
offering a great advantage compared with Haar-random 
circuits. This enables comprehensive studies of 
universal phenomena including information scrambling, quantum chaos, etc., 
that are often inaccessible in conventional condensed-matter 
systems. 
The system is also versatile since they can be straightforwardly extended to 
incorporate other ingredients, 
such as long-range gates~\cite{Nahum:2021aa,Block:2022aa,Sharma:2022aa} 
or multi-qubit projective measurements~\cite{Lavasani:2021aa,Lavasani:2021ab},
spacetime duality~\cite{Ippoliti:2021aa,Lu:2021aa,Ippoliti:2022aa},
higher dimensions~\cite{Turkeshi:2020aa,Sierant:2022aa,Liu:2022aa,Feng:2023aa,Wei:2026aa}, etc.
Most notably, this system demonstrates a 
measurement-induced phase transitions (MIPT) between entanglement
volume-law and area law phases~\cite{Li:2018aa,Chan:2019aa,Skinner:2019aa,Szyniszewski:2019aa}, 
arise from the
competition between unitary evolution and projective measurements.
Beyond the phase diagram, MIPT has been connected to the
transition to dynamical purification and emergent quantum error correction
\cite{Choi2020QEC,Gullans:2020aa,Li:2021aa}. 
Clifford circuits have also provided a controlled setting for resolving
conformal criticality and the geometry of entanglement clusters
\cite{Li2021Conformal,Lunt2021Clusters}.
Meanwhile, related measurement-only and symmetry-constrained hybrid circuits
have shown that measurements can
also generate or protect symmetry-breaking and topological phases
\cite{Ippoliti2021MeasurementOnly,Lavasani2021Topological,
Sang2021MeasurementProtected,MorralYepes2023SPT}.
Alongside these developments, Clifford-based platforms have opened new
directions involving multipartite entanglement, magic and classical
simulability, and measurement-induced quantum advantage
\cite{Lira-Solanilla:2025aa,Sharma2026Fractal,Fux2024Nonstabilizerness,
Bejan2024Magic,Scocco2026Nonstabilizerness,
Watts2025QuantumAdvantage}.
As such, monitored random Clifford circuits have emerged as a fertile ground both 
for revisiting longstanding questions and for uncovering novel physics. 

Much of the difficulty in the field stems from the fact that, 
while a direct understanding of the output states would provide a complete 
physical picture, the available analytical tools have been designed 
exclusively to compute bipartite entanglement, thus offering only a partial 
view of the output ensemble.
Below we briefly review two established theoretical formalisms 
and discuss their applications in the context of tripartite
Greenberger--Horne--Zeilinger (GHZ) entanglement, illustrating
their restricted applicability.

The first formalism uses the replica trick to evaluate the 
averaged $n$th R\'{e}nyi entropy 
$\langle S_n\rangle =-\frac{1}{n-1}\langle \ln Z_n \rangle$,
where $\langle\cdot\rangle \equiv \mathrm{E}_U[\cdot]$ 
(assuming the circuit is unitary). Nonlinearity of this formula comes from
two sources, one is the logarithm $\ln(Z_n)$. 
A replica index $m$ is introduced so that 
$\ln(Z_n)=\lim_{m\rightarrow 0}[(Z_n)^m-1]/m$. The other
nonlinearity arises from $Z_n=\tr(\rho_A^n)$, where 
$\rho_A=\mathrm{tr}_{\bar{A}}\left[U \rho_{A,\bar{A}}(0) U^\dagger\right]$ 
is the reduced density matrix of subsystem $A$ obtained
from some full-system initial state $\rho_{A,\bar{A}}(0)$.
We need another replica index $n$ so that 
$\tr(\rho_A^n)$ is replaced by $\tr(\rho_A^{\otimes n} \hat{\Sigma})$,
where $\hat{\Sigma}$ is an operator fixing the contraction rule 
among the $n$ replica. The averaged R\'{e}nyi entropy then takes
the form $\mathrm{E}_U \left[U^{\otimes nm}(\cdot)(U^\dagger)^{\otimes nm}\right]$. Then, the Schur--Weyl duality and the Weingarten calculus~\cite{Weingarten} reduces
the formula to a network of classical variables, one for each gate~\cite{Bao:2020aa,Jian:2020aa,Zhou:2019aa}. 
Notably, each variable takes value from the permutation group
$S_{nm}$ when $nm\leq 3$, but from
a larger set when $nm> 3$~\cite{Gross:2021aa,Li:2024aa} because 
the Clifford group forms only a unitary 3-design rather than
a fully Haar-random ensemble~\cite{Zhu:2017}. The required analytical
continuation in the replica index is generally intractable.

The second approach is based on the clipped gauge~\cite{Nahum2017PRX},
a rule for selecting the stabilizer generators of the stabilizer 
state generated by the Clifford circuits.
In a one-dimensional (1D) chain, let $\rho_{l(r)}(x)$
denotes the number of stabilizer generators whose
left(right) endpoint is located at position $x$. The clipped gauge
imposes the local constraint $\rho_{l}(x)+\rho_r(x)=2$, $\forall x$.
Taking $\rho_r(x)$ as the independent
variable, the entanglement entropy of the subsystem to 
the right of the cut at $x$ can be expressed as $S_x=\sum_{y>x}[\rho_r(y)-1]$. 
Since $\rho_r(x)\leq 2$, the evolution of $\rho(x)$,
in the continuous time and long-wavelength limit, fits into the noisy Burgers 
equation~\cite{Halpin-Healy:1995aa}. As a consequently, the
entanglement entropy $S_x$
satisfies the Kardar--Parisi--Zhang (KPZ) 
equation describing directed polymer in a 
random environment~\cite{Kardar:1986aa}. It explains the
linear growth of entanglement entropy and the 
$t^{1/3}$-subleading fluctuation~\cite{Fisher:2023aa}. But it seems
not easy to incorporate measurements systematically~\cite{Li:2021aa}.

As is clear from the above, both approaches target the 
computation of entropy. But how about other
facets of the correlation structure that 
quantum information theory has illuminated?
We notice that recently irreducible multipartite entanglement has 
attracted growing interest in both many-body 
physics~\cite{Nezami:2020aa,Siva:2022aa,KanePRX2022,Liu:2022aa,Liu:2024ab,BergPRL2025,ZouYijian2021PRL,LiuKe2026Arxiv,Zhou:2026aa}, high energy 
physics~\cite{Akers2020JHEP,Hayden2021JHEP,Balasubramanian2025JHEP,Iizuka2025JHEP,SharmaPRD2022,Chris2024,iizuka2025,iizuka2025b,SimonPRL2026}.
Clifford circuits likewise offer a natural testbed for 
investigating multipartite GHZ entanglement~\cite{Fattal:2004aa,Bravyi:2006aa,Xu2025prb}.

\subsection{Challenges Raised by GHZ Entanglement}
A theorem proved in Ref.~\cite{Bravyi:2006aa} states
that any tripartite stabilizer state $\ket{\Psi}_{ABC}$ 
can be decomposed into 
a collection of three-qubit GHZ state $\ket{\mathrm{GHZ}}=(\ket{000}+\ket{111})/\sqrt{2}$, two-qubit 
Bell state $\ket{\psi}=(\ket{00}+\ket{11})/\sqrt{2}$, and 
single qubit, by party-local 
Clifford (PLC) unitaries defined in the form of 
$U_A\otimes U_B\otimes U_C$:
\begin{equation}
    \label{eq:decomposition}
    \begin{split}
    \ket{\Psi}_{ABC}\xrightarrow{\text{PLC}} & \ket{\psi_{AB}}^{\otimes n_{AB}} \ket{\psi_{BC}}^{\otimes n_{BC}}
    \ket{\psi_{AC}}^{\otimes n_{AC}} \\
        &\otimes \ket{\mathrm{GHZ}}^{\otimes g_3}
    \end{split}
\end{equation}
where the single qubits are not shown. The index $g_3$ counts
the number of $\ket{\mathrm{GHZ}}$ that can be extracted 
from $\ket{\Psi}$ by PLC. It is further proved that 
\begin{equation}
    \label{eq:bravyi_g3}
    g_3=N-\dim \mathcal{S}_{\text{local}}
\end{equation}
where $N$ is the number of qubits, and 
$\mathcal{S}_{\text{local}}$ is the group generated by
stabilizers which are trivial in at least one subsystem~\cite{Fattal:2004aa,Bravyi:2006aa}.

\begin{figure}
    \centering
    \includegraphics[width=0.95\linewidth]{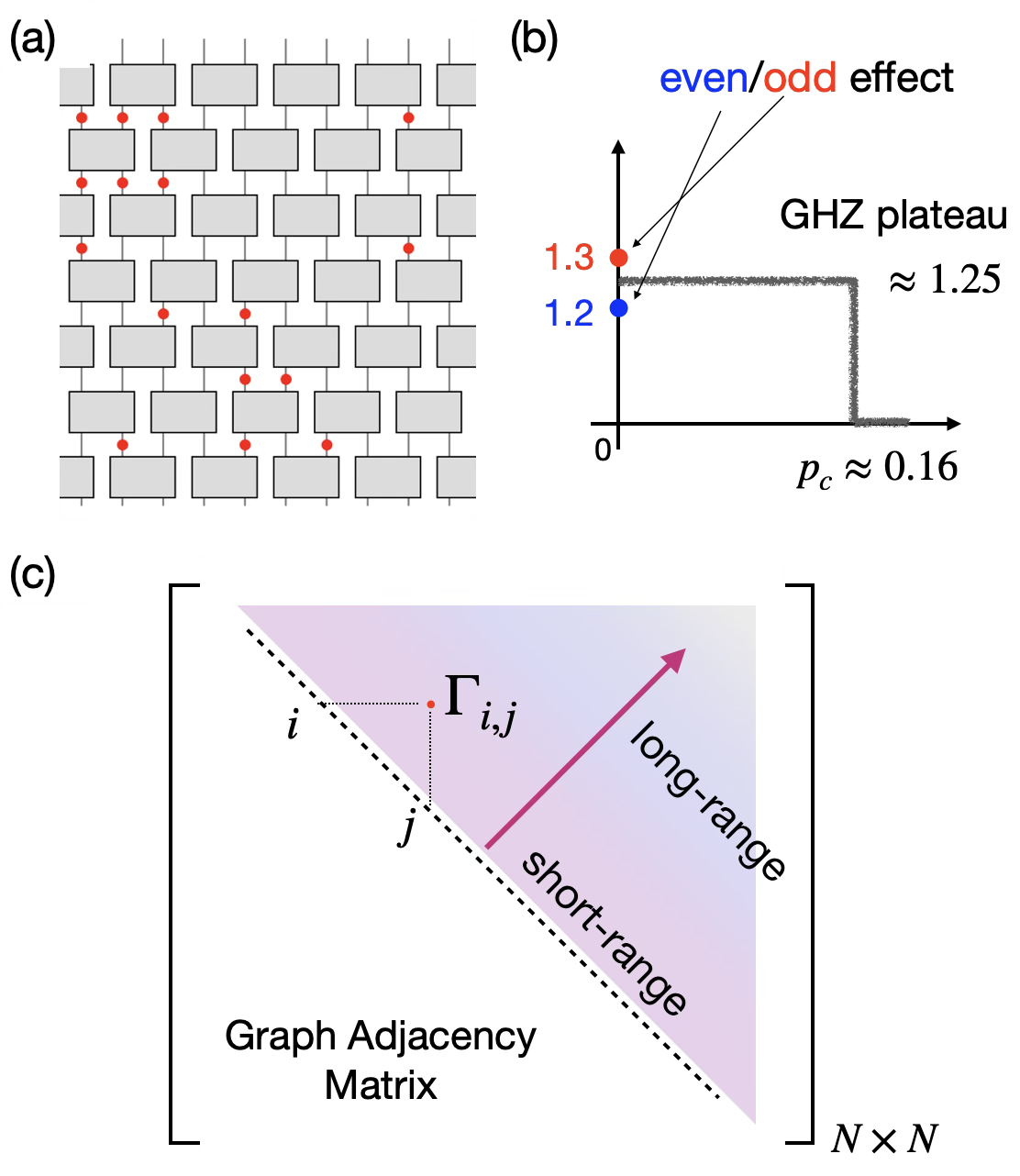}
    \caption{(a) Brickwork random Clifford circuits consists of local two-qubit Clifford gates interleaved with
    random single-qubit Pauli $Z$ measurements (red dots). (b) Schematic
    of the GHZ-entanglement content as a function of the measurement rate $p$ 
    in monitored circuits: An even--odd 
    effect at $p=0$ and a plateau $\langle g_3\rangle\approx 1.25$
    in bulk of the volume-law phase. (c) Our primary theoretical tool is the graph
    adjacency matrix ($\Gamma$) for the output stabilizer states. 
    In the large-$N$ limit, the statistics of such adjacency matrix
    determines the distribution of stabilizer states sampled by the
    monitored random Clifford circuits. Due to translational invariance in the bulk of the system, matrix elements $\Gamma_{i,j}$ with the same $d=\abs{i-j}$ is statistically homogeneous. Larger $d$ corresponds to 
    longer-range connections in the graph (under open boundary condition).}
    \label{fig:system}
\end{figure}

An immediate question is the expected GHZ content 
$\langle g_3\rangle$. This was first studied in
Ref.~\cite{Smith:2006aa} but Eq.~(5) thereof contains an error. 
Recently, it was proved in
Ref.~\cite{Nezami:2020aa} that the GHZ content of random stabilizer network
is $O(1)$. One of the author and
colleague have numerically simulated unitary brickwork Clifford circuits
and found an even--odd effect: $\langle g_3\rangle\approx 1.2$ and 
$1.3$ for even and odd $N$, respectively, provided the
sizes of the three subsystems satisfy the triangle inequalities~\cite{Xu2025prb}. 
More interestingly, as depicted 
in Fig.~\ref{fig:system}(b), $\langle g_3\rangle$ remains at a
plateau of approximately $1.25$, middle point of the
even--odd effect, throughout the volume-law phase. 
Similar plateaus against measurement rate $p$ 
are also seen in the standard deviation of $g_3$.
So, what is the fundamental reason underlying such
insensitivity to $p$.

To address it by the established methods, it was noticed
that 
Eq.~\eqref{eq:decomposition} implies an
alternative formula for $g_3$~\cite{Sang:2021aa,Bertini:2022aa}
in terms of the quantum mutual information $\mathcal{I}_{A:B}$ and
the quantum negativity $\mathcal{N}_{A:B}$~\cite{Vidal:2002aa}:
\begin{equation}
    g_3= \mathcal{I}_{A:B}-2\mathcal{N}_{A:B}.
\end{equation}
The same result holds for any
other choice of subsystem pair. Both terms can be evaluated by the
replica trick~\cite{Calabrese2005,Calabrese:2012aa}.
One strategy is thus to calculate each of them, and find the gap. 
However, both $\mathcal{I}_{A:B}$ and $\mathcal{N}_{A:B}$ diverge linearly 
with $N$, and their sub-leading corrections include terms of order $N^{1/3}$~\cite{Li:2023aa,Weinstein:2022aa} and perhaps
$\log(N)$, whereas their gap $\langle g_3\rangle$ is an O(1) constant. 
It is unlikely that KPZ equation is quantitatively precise to 
$O(1)$ level. Indeed, theoretically one would obtain $\mathcal{I}_{A:B}=2\mathcal{N}_{A:B}$~\cite{Sang:2021aa,Weinstein:2022aa},
although numerically a finite gap is visible~\cite{Sang:2021aa}.

The crux of the matter is that the established methods, 
rather than targeting the output states directly, 
are designed to compute bipartite entanglement probes. But it is 
the former that is conceptually primary. This motivates us to
develop a new theoretical formalism.

\subsection{What Is New in This Article}

Our theoretical formalism is built upon graph state~\cite{Schlingemann:2001aa,Raussendorf:2003aa},
a special kind of stabilizer state whose stabilizer
generators admit a graph representation, see Sec.~\ref{sec:preliminary}. 
Such graph representation can be extended to a general
stabilizer state up to local Clifford unitaries 
(which means product of single-qubit Clifford gates).
Since the output ensemble of the random circuits is
invariant against local Clifford unitaries,
to study the output states is equivalent to study their
graph representations.

Along this line, the first subtlety we need to address is
that the graph representation is not unique: Two graphs different
by a sequence of 
\emph{local complementations}~\cite{Kotzig:1968aa} 
correspond to local-Clifford-equivalent states.
Recently, a one-to-one correspondence is established
between stabilizer states and graph states extended by
a restricted set of local Clifford gates~\cite{Hu:2022aa}. 
In this gauge, we observe that the graph ensemble, with the
multiplicity accounted, 
approaches asymptotically the ensemble of Erd\H{o}s--R\'{e}nyi random graph
$G(N, 1/2)$, i.e., $N$-vertex graphs where each edge appears with
probability $1/2$~\cite{Erdos:1960aa,Gilbert:1959aa},
see Sec.~\ref{sec:key_observation}. This allows us to derive
the even--odd effect of GHZ content analytically in Sec.~\ref{subsec:<g3>},
demonstrating the power of the graph-based framework.

Entries of the graph adjacency matrix are actually the
effective dynamical variables.
As indicated in Fig.~\ref{fig:system}(c), quasi-translation invariance
implies the dependence of $\langle \Gamma_{i,j}\rangle$ 
on distance $d=\abs{i-j}$. Thus, the short- and
long-range properties are encoded in entries 
separated geometrically, allowing
us to extract local and nonlocal behaviors conveniently.
In Sec.~\ref{sec:AED}, we uncover a homogeneous long-range averaged
edge density (AED): $\langle\Gamma_{d}(p)\rangle\approx (1-p/p_c)/2$,
where $p_c$ is the MIPT critical point. Notably it is flat against
$d$ if $d$ is sufficiently large. This is another evidence
of a hitherto hidden structure in the
output states in the volume-law phase.

This structure is identified as a subgraph with a uniform
edge probability $1/2$. To reveal it, we translate the gates 
and measurements into
graph operations and identify 
three types of edge toggles denoted by $C_\ell$ with $\ell=0,1,2$,
corresponding to different levels of nonlocality.
The most nonlocal $C_2$ toggles induce 
an subgraph in the class of Erd\H{o}s--R\'{e}nyi $G(N_{\sub}, 1/2)$,
where the relative size $N_{\sub}/N\approx\sqrt{1-p/p_c}$, see Sec.~\ref{subsec:C2toggle}.
Crucially, the edge probability $1/2$ holds for all $p$. Such 
random subgraph corresponds to a unitary random Clifford 
circuit on $N_{\sub}$ qubits. The graph adjacency matrix is
much sparser in other blocks. Thus, an output state of monitored
circuits is equivalent to the output of an $N_{\sub}$-qubit unitary circuit,
weekly perturbed by $N-N_{\sub}$ qubits that carry little entanglement. 
From this physical picture,
the GHZ plateaus, along with the quantum error-correcting capability
of the volume-law phase,
can be understood directly as arising from the ``hidden'' unitary circuit.
We further reveal a clustering effect of the
dense vertices
by simplifying the circuits dynamics into
an infection-recovery toy model, see Ref.~\ref{subsec:toymodel}. 
This toy model also exhibits an
absorbing-state phase transition 
which gives MIPT another interpretation.

At last, we show the $C_1$ toggles can be viewed as a local interaction between
the entries of the graph adjacency matrix. Based on this observation, 
we propose a mean-field argument in Sec.~\ref{subsec:pc}, 
which determines the  
critical point of MIPT at $p=0.1608$, in remarkable agreement with the
numerical result $p_c\approx 0.16$. This further underscores 
the strength of our graph-theoretic approach, as such an
theoretical prediction has so far eluded all previous treatments.

To summarize, our framework goes beyond bipartite quantum correlation and
offers a more versatile and powerful analytical toolkit, 
capable of addressing problems that have so far 
resisted theoretical understanding. In Sec.~\ref{sec:conclusion} we
discuss possible future developments of our method and
potential applications.

\section{Preliminaries: Systems, Stabilizer States, and Graph States}
\label{sec:preliminary}

\subsection{Stabilizer States and Clifford Circuits} 

We refer the readers to the textbook of Nielsen and Chuang~\cite{Nielsen:2010aa} 
for a systematical introduction of stabilizer 
states. Here we just present the minimal.
An $N$-qubit stabilizer state is the
simultaneous eigenstate (with eigenvalue $+1$) of $N$ independent
and mutually commuting stabilizer generators. These
generators are Pauli strings, a tensor product of 
Pauli operators ($X,Y,Z$) and the identity operator ($\mathbb{I}$)
up to an overall sign $\pm 1$ or $\pm i$.
Since $Y\propto XZ$,  a Pauli string can be expressed 
as $\otimes_{i=1}^{n} X_{i}^{x_i}Z_{i}^{z_i}$, where
$x_i, z_i \in\{0,1 \}$. When we say a stabilizer generator acts trivially
on the $i$th qubit, we mean $x_i=z_i=0$. By this notation,
this Pauli string is specified by a $2N$-binary string $(\vec{x},\vec{z})$. 
Collecting the binary strings of all the $N$
stabilizer generators, one obtains an $N\times 2N$ matrix
$\left(\mathcal{S}_X\vert \mathcal{S}_Z\right)$,
which is called the binary symplectic representation.
We shall henceforth abbreviate ``stabilizer generator'' as 
``stabilizer'' when no confusion arises.

The system we study is the 1D monitored random Clifford circuit in the brickwork layout
illustrated in Fig.~\ref{fig:system}(a). Therein, each
block represents an independent random two-qubit Clifford gate.
After each gate, every qubit is measured along the basis of 
Pauli $Z$ operator with probability $p$. A trajectory of stabilizer state
is thus determined by the realizations of each gate, the locations of
measurements, and the measurement outcomes. 
One can calculate anything 
from every individual sample of state trajectory. The statistics of this quantity 
among all realizations of the random circuits is what people 
calculate
in this system. We are motivated by a calculation of
the GHZ entanglement that can be extracted from the output states~\cite{Xu2025prb}.

\subsection{Graph State}

Graph is a mathematical concept defined by a set of vertices 
and a set of edges connecting pairs of vertices. A graph is specified
by its adjacency matrix (denoted by $\Gamma$ throughout this Article),
which is a symmetric matrix with zero diagonals. 
An entry $\Gamma_{i,j}$ equals $1$ 
if the two vertices $v_i$ and $v_j$
are connected by an edge. Otherwise $\Gamma_{i,j}=0$.
Given a graph $\Gamma$ with $N$ vertices, a graph state $\ket{\Gamma}$
is defined by firstly putting one qubit initialized in $\ket{+}$
(the $+1$ eigenstate of Pauli $X$) on every vertex,
and secondly implementing  $\mathrm{CZ}$ gate on pairs of qubits connected by 
an edge~\cite{Schlingemann:2001aa,Raussendorf:2003aa}. The $\mathrm{CZ}$ gate is 
symmetric with respect to (w.r.t.) the relevant qubits
\begin{equation}
\label{eq:cz}
    \mathrm{CZ}=\frac{1}{2}\left(
    \mathbb{I}\otimes\mathbb{I}+\mathbb{I}\otimes Z+
    Z\otimes\mathbb{I}-Z\otimes Z
    \right).
\end{equation}
The prepared state $\ket{\Gamma}$ is written as
\begin{equation}
    \ket{\Gamma}=\left(\prod_{i<j}\mathrm{CZ}_{i,j}^{\Gamma_{i,j}}\right)
    \ket{+}^{\otimes N}.
\end{equation}
This is a stabilizer state whose stabilizers 
can be directly readout from the graph:
Every vertex $v_i$ contributes an independent stabilizer 
\begin{equation}
    \label{eq:graph-stabilizer}
    \mathcal{S}_i= X_i\bigotimes_{j\in\mu(i)}Z_j=
    X_i\bigotimes_{j=1}^{N} Z_j^{\Gamma_{i,j}}
\end{equation}
where $\mu(i)$ denotes the \emph{neighbors} of $v_i$, i.e.,
vertices connected to $v_i$ (do not confuse it with
the nearest-neighbors in the 1D chain).

To relate the graph representation to the standard binary
symplectic representation
of the stabilizers
$\left(\mathcal{S}_X\vert \mathcal{S}_Z\right)$,
we can manipulate the latter by freely swapping the 
rows, adding one row to another (module 2), and switching 
columns of this matrix. These operations 
do not change the Abel stabilizer group.
If $\mathcal{S}_X$ is full rank, the above operation can implement
Gaussian elimination that leads to an
equivalent representation
\begin{equation}
    \label{eq:def-graph}
    \left(\mathcal{S}_X\vert \mathcal{S}_Z\right) \Leftrightarrow \left(\,\mathbb{I}_{N}\, \big\vert \, \Gamma \,\right)
\end{equation}
where $\mathbb{I}_N$ is the $N$ dimensional identity matrix and $\Gamma$
has a form of adjacency matrix (ensured by
the defining condition of the stabilizers that $[\mathcal{S}_i, \mathcal{S}_j]=0$, $\forall i,j$).

If $\mathcal{S}_X$ is not full rank, the stabilizer state itself is 
not a graph state.
But we can always make $\mathcal{S}_X$ full rank by applying 
local Clifford gates,
since the matrix $\left(\mathcal{S}_X\vert \mathcal{S}_Z\right)$ has full row rank.
In the end, any stabilizer state $\ket{\psi}$ is 
local-Clifford-equivalent to a graph state:
\begin{equation}\label{eq:extended-graph-state}
   \ket{\psi}= \left(\bigotimes_{i=1}^{N}U_i\right)\,\ket{\Gamma}.
\end{equation}
One might therefore expect that every graph 
defines a local-Clifford-equivalent
class of stabilizer states. This is not true because 
two graph states are different by 
single-qubit Clifford groups if and only if their graphs can be converted to
each other via a sequence of \emph{local complementations}~\cite{Van-den-Nest:2004aa}. Local complementation
at a vertex $v_1$ is the graph operation corresponding to
the mapping that for any pair
$v_i, v_j \in \mu(1)$, $\Gamma_{i,j}\rightarrow \Gamma_{i,j}+ 1\;
(\mathrm{mod}\;{2})$~\cite{Kotzig:1968aa}.
Denote the new graph state by $\ket{v_1(\Gamma)}$. 
It relates with the original graph state by 
\begin{equation}
\label{eq:local complementation}
    \ket{v_1(\Gamma)}=e^{-i\frac{\pi}{4}X_1}\prod_{j\in \mu(1)} e^{i\frac{\pi}{4}Z_j}\ket{\Gamma},
\end{equation}
Or equivalently, 
\begin{equation}
  \ket{\Gamma}  =H_1 S_1^\dagger H_1 \prod_{i\in\mu(1)} S_i \ket{v_1(\Gamma)}
\end{equation}
where $H=\frac{1}{\sqrt{2}}
\begin{pmatrix}
    1 & 1 \\
    1 & -1
\end{pmatrix}$ and $S=\begin{pmatrix}
    1 & 0 \\
    0 & i
\end{pmatrix}$.
Determining weather two graphs are equivalent up to local complementations 
requires an algorithm with complexity 
$O(N^4)$~\cite{Van-den-Nest:2004ab}.

\subsection{Canonical extended graph state}
\label{subsec:canonical_graph}

Recently, Hu and Khesin~\cite{Hu:2022aa} proposed a gauge on the 
right hand side of Eq.~\eqref{eq:extended-graph-state}, restricting it to
\begin{equation}
\label{eq:canonical_graph}
\ket{\psi}=\left(\bigotimes_{i=1}^N \hat{c}_i \hat{z}_i\right)\ket{\Gamma},
\end{equation}
where $\hat{c}_i\in \{\mathbb{I}, S, H\}$ and
$\hat{z}_i\in \{\mathbb{I}, Z\}$.  Moreover, 
based on a given numbering of the qubits, it is required that
when $\Gamma_{i,j}=1$, either $c_i\neq H$ or 
$c_j\neq H$; meanwhile, if $c_i=H$ there must be $j>i$. 
States~\eqref{eq:canonical_graph} satisfying these conditions are called \emph{canonical extended graph states}.
They are in one-to-one correspondence with the stabilizer states.

\section{GHZ Statistics: The Even--Odd effect}
\label{sec:nullity}

In this section, we develop the idea of reducing random 
stabilizer states to random graphs and use it to derive
the GHZ statistics, especially the even--odd effect.
Random stabilizer states are also the output ensemble of random Clifford circuits 
(with measurements) in the
1D brickwork layout with depth $O(N)$~\cite{Brandao:2016aa}.
Hereafter we shall use ``vertex'' and ``qubit'' interchangeably  
without any risk of ambiguity.

\subsection{From Random Stabilizer States to Random Graphs}
\label{sec:key_observation}

We review the counting argument proving the
one-to-one correspondence between stabilizer states and
canonical extended graph state~\cite{Hu:2022aa}. 
Consider an iterative construction of the graph by sequentially adding 
vertices (hence, a numbering of vertices is introduced automatically). 
When the $k$th vertex $v_k$ is included,
the canonical form dictates that if $\hat{c}_k=H$ 
no edges connect $v_k$ to any preceding vertices $v_i$ ($i<k$). 
Conversely, if 
$\hat{c}_k\neq H$, then $v_k$ can be connected to an arbitrary 
subset of $\{v_i\}_{i<k}$. Enumerating the possibilities, the isolated 
case ($v_k$ has no edges) admits 6 choices for  $\hat{c}_k\hat{z}_k$;
the connected case admits 4 choices for $\hat{c}_k\hat{z}_k$ and 
$2^{k-1}-1$ ways to choose at least one edge to the preceding vertices.
The total number of new configurations brought by 
adding the $k$th vertex is therefore 
$n_k=6+4(2^{k-1}-1)=2(2^{k}+1)$. 
Taking the product over all vertices yields
\begin{equation}
\prod_{k=1}^N n_k = 2^N \prod_{k=1}^N (2^k+1).
\end{equation}
This formula gives the total number of $N$-qubit 
canonical extended graph state, which exactly 
matches the total number of $N$-qubit stabilizer state.

The distribution of the graph representations can be obtained by 
counting the multiplicity of each graph. For this purpose, it is more convenient 
to consider the probability that a pair of vertices ($v_{i}, v_k$) with $i<k$ is connected by an edge:
\begin{equation}
\label{eq:key}
    P_{i<k}=\frac{2^k}{2(2^k+1)}\approx \frac{1}{2}(1-2^{-k})
    \xrightarrow{\text{large}\; k} \frac{1}{2}.
\end{equation}
This is based on the counting that when $v_k$ is introduced,
among the $n_k=2(2^k+1)$ total configurations, exactly
$2^k$ (i.e., $4\times 2^{k-2}$) have $v_i$ and $v_k$
connected, and the action of further adding vertices $\{v_j\}_{j>k}$ 
neither depends on the subgraph induced by $\{v_{i}\}_{i\leq k}$
nor changes it. Thus, we learn from
Eq.~\eqref{eq:key} that the ratio of 
edges of which the probability to appear is less than $1/2-\epsilon$ is 
about $\left[\log_2(1/\epsilon)/N\right]^2$, for any small positive constant 
$\epsilon$. This number becomes negligible in the large
$N$ limit. Therefore, for practical purposes we can view it
as Erd\H{o}s--R\'{e}nyi random graph
$G(N,r=1/2)$, the ensemble of graphs where the probability for an
edge to present is homogeneously $r$~\cite{Erdos:1960aa,Gilbert:1959aa}.

We note an additional source of gauge freedom 
associated with the numbering of vertices. Imagine that every vertex has a ``name''
and a number. For a given graph (where every edge is specified by the names of
the vertices),
different vertex numbering can lead to different numbers of canonical extended graph 
states. By averaging over all numberings, 
the resulting distribution over graphs becomes more uniform, 
approaching closer the Erd\H{o}s--R\'{e}nyi random graph
$G(N,r=1/2)$.

\subsection{Express $g_3$ in Graph Notation}
\label{subsec:g3}

Partition the rows and columns of the adjacency matrix $\Gamma$ according to the vertex subsets $A,B,C$:
\begin{equation}
    \label{eq:gamma}
    \Gamma\equiv\begin{bmatrix}
    \Gamma_A \\
    \Gamma_B \\
    \Gamma_C
    \end{bmatrix}=\begin{bmatrix}
        \Gamma_{AA} & \Gamma_{AB} & \Gamma_{AC} \\
        \Gamma_{AB}^{\mathrm{T}} & \Gamma_{BB} & \Gamma_{BC} \\
        \Gamma_{AC}^{\mathrm{T}} & \Gamma_{BC}^{\mathrm{T}} & \Gamma_{CC}
    \end{bmatrix},
\end{equation}
where $\mathrm{T}$ denotes matrix transpose.
We find that theorem~\eqref{eq:bravyi_g3} 
can be rephrased as 
\begin{equation}
    \label{eq:g3}
g_3=\dim\mathrm{ker}\, \Gamma -\sum_{\omega=A,B,C}\dim\mathrm{ker}\,\Gamma_\omega,
\end{equation}
where $\dim\ker(\cdot)$ denotes the nullity (dimension of the kernel)
over $\mathbb{F}_2$. One can substitute the decomposed state~\eqref{eq:decomposition} into the right hand side of
Eq.~\eqref{eq:g3} to verify its correctness.
Explicitly, an isolated qubit corresponds to a zero on the diagonal.
A Bell pair is equivalent to $\frac{1}{\sqrt{2}}\left(|0,+\rangle+|1,-\rangle \right)$, where $X\ket{\pm}=\pm \ket{\pm}$. The corresponding
adjacency matrix is a block $\begin{pmatrix} 0 & 1 \\ 1 & 0 \end{pmatrix}$.
A GHZ state is equivalent to 
$\frac{1}{\sqrt{2}}\left(|+,0,+\rangle+|-,1,-\rangle \right)$,
whose adjacency matrix reads 
$\begin{pmatrix} 0 & 1 & 0 \\ 1 & 0 & 1 \\ 0 & 1 & 0 \end{pmatrix}$.
The graph adjacency matrix of the decomposed state~\eqref{eq:decomposition} is
thus a diagonal of such blocks. One thus obtains
\begin{equation}\label{eq:nullity-A}
\dim\ker\,\Gamma_A=N_A-(n_{AB}+n_{AC}+g_3)
\end{equation}
where $N_A$ is the number of qubits in $A$.
Analogous formulae can be obtained for $B$ and $C$. 
For the whole graph we have
\begin{equation}
\dim\ker\,\Gamma=N-2(n_{AB}+n_{AC}+n_{BC})-2g_3.
\end{equation}
The validity of Eq.~\eqref{eq:g3} follows straightforwardly.

Then, we need to show Eq.~\eqref{eq:g3} is PLU-invariant.
Write $\Gamma_A=[\Gamma_{AA}, \Gamma'_{A}]$ with $\Gamma'_{A}=[\Gamma_{AB},\Gamma_{AC}]$. 
The decomposition~\eqref{eq:decomposition} has $\Gamma_{AA}=0$.
The rank of $\Gamma'_A$, $n_{AB}+n_{AC}+g_3$, is PLU 
invariant (hence also the nullity) because it encodes the entanglement
between $A$ and its complement. 
The kernel of $\Gamma'_A$ consists of binary vectors $\vec{v}\in\mathbb{F}_2^{N_A}$ 
satisfying $\vec{v}\cdot\Gamma'_A=0$. This vector may not in $\ker \Gamma_{AA}$,
rendering $\dim \ker\Gamma_A<\dim\ker\Gamma'_A$. But 
simultaneously this vector is also subtracted from the kernel of the full matrix $\Gamma$. 
This is why Eq.~\eqref{eq:g3} is generally true.

\subsubsection{A more rigorous treatment}
The above argument is heuristic but captures the essential physics.
Here we present another proof having more sense of rigor. 

Due to Eq.~\eqref{eq:def-graph}, the binary symplectic 
representation of the 
stabilizer group of a graph state is
\begin{equation}
\mathcal{S} = \{\, (x,\,\Gamma x) \mid x\in\mathbb{F}_2^N \,\}.
\end{equation}
A stabilizer is \emph{not supported} on party $\omega\in\{A, B, C\}$ 
if and only if $x_\omega = 0$ and $(\Gamma x)_\omega = 0$, where
the subscript means projection onto $\omega$.
Define
\begin{equation}
V_\omega = \{\, x\in\mathbb{F}_2^N \mid x_\omega = 0,\; (\Gamma x)_\omega = 0 \,\}.
\end{equation}
The subspace spanned by all stabilizers that miss at least one party is $V_A+V_B+V_C$, where ``+'' denotes the sum of subspaces.  Then,
Eq.~\eqref{eq:bravyi_g3} is rephrased to
\begin{equation}
g_3 = N - \dim\, (V_A+V_B+V_C). 
\label{eq:standard}
\end{equation}
Let $W_\omega = \operatorname{span}\{e_i : i\in\omega\}$ be the coordinate subspace of party~$\omega$; hence $\mathbb{F}_2^N = W_A\oplus W_B\oplus W_C$.  
The conditions $x_\omega = 0$ and $(\Gamma x)_\omega = 0$ are equivalent to $x \in W_\omega^\perp$ and $x \in (\Gamma W_\omega)^\perp$, therefore
\begin{equation}
V_\omega = (W_\omega + \Gamma W_\omega)^\perp .
\end{equation}
Consequently
\begin{equation}
V_A+V_B+V_C = \bigl[ \bigcap_{\omega \in\{A,B,C\}}\bigl(W_\omega+\Gamma W_\omega\bigr) \bigr]^\perp ,
\end{equation}
and~\eqref{eq:standard} becomes
\begin{equation}
g_3 = \dim\, \bigcap_{\omega\in\{A,B,C\}} \left(W_\omega + \Gamma W_\omega\right). \label{eq:intersection}
\end{equation}

Observe that for any $v\in W_\omega$ the intra‑party component $(\Gamma v)_\omega$ again lies in $W_\omega$.  
It can therefore be absorbed into the $W_\omega$ summand without changing the subspace $W_\omega+\Gamma W_\omega$.  
Explicitly, define the \emph{inter‑party} adjacency matrix
\begin{equation}
    \label{eq:gamma}
    \Gamma_{\text{off}}\equiv \begin{bmatrix}
        0 & \Gamma_{AB} & \Gamma_{AC} \\
        \Gamma_{AB}^{\mathrm{T}} & 0 & \Gamma_{BC} \\
        \Gamma_{AC}^{\mathrm{T}} & \Gamma_{BC}^{\mathrm{T}} & 0
    \end{bmatrix}.
\end{equation}
Then we have
\begin{equation}
W_\omega + \Gamma W_\omega = W_\omega + \Gamma_{\text{off}} W_\omega,
\end{equation}
which means $\Gamma$ can be safely replaced by $\Gamma_{\text{off}}$.
Then, the substitution of Eq.~\eqref{eq:decomposition} is justified.

\subsection{Express $\langle g_3\rangle$ in Graph Notation }

To evaluate the expectation value $\langle g_3\rangle$, 
we first consider the
$\langle \dim\ker\Gamma_{\omega} \rangle$ with $\omega=A$ for example.
In the decomposition $\Gamma_A=[\Gamma_{AA}, \Gamma'_{A}]$,
$\Gamma_{AA}$ is the adjacency matrix of a subgraph. Hence, it is restricted
to be symmetric, zero-diagonal matrix. However,
$\Gamma'_A$ is free from such restriction. Every entry of it 
is an independent binary randomness.

$\Gamma'_A$ is an $N_A \times (N_B+N_C)$ matrix.
If $N_A> N_B+N_C$, $\Gamma'_A$ has more rows than columns,
thus, it never has full row rank. On the other hand, if the triangle inequality is satisfied, i.e.,
$N_A< N_B+N_C$, 
the probability that $\Gamma'_A$ attains full row rank reads
\begin{equation}
\label{eq:rank-A}
P_A=  \prod_{i=0}^{N_A-1}\left[1-2^{i-(N_B+N_C)} \right]
\end{equation}
It is obtained by counting the number of options for adding 
a new row linear independent to all preceding rows simultaneously.

We define ``thermodynamic limit'' by diverging $N$ with
a fixed ratio $N_A{:}N_B{:}N_C$.  
Then Eq.~\eqref{eq:rank-A} approaches 1 exponentially fast. Consequently, 
$\langle \dim\ker \Gamma_A\rangle = 0$ almost surely, consistent 
with standard perspective that after a deep circuit all qubits in $A$
are entangled with qubits outside of $A$, 
if $A$ contains no more than half of all qubits~\cite{Nahum2017PRX}.
The same reasoning also works for $\langle \dim\ker \Gamma_B\rangle$ 
and $\langle \dim\ker \Gamma_C\rangle$.
Therefore, when the sizes of the subsystems 
$(N_A, N_B, N_C)$ fulfills the triangle inequalities, 
the GHZ entanglement can be extracted, $\langle g_3\rangle$, 
is reduced to intrinsic graph properties independent to the partitioning,
\begin{equation}
\label{eq:aveg3}
\langle g_3\rangle =\langle \dim\ker \Gamma \rangle.
\end{equation}
As an additional remark,
numerical calculations found that the plateau of GHZ entanglement
drops to zero via a partitioning-induced phase transition
if the triangle inequalities are disobeyed~\cite{Xu2025prb}. The vanishing of 
$g_3$ in this case can also be proved by using Jensen's inequality based on Eq.~\eqref{eq:decomposition}~\cite{Xu2025prb,Nezami:2020aa}.

\subsection{Deriving the Even--Odd Effect}
\label{subsec:<g3>}

Here we derive $\langle g_3\rangle$ through the
distribution of $g_3$ over graph states corresponding to the
Erd\H{o}s--R\'{e}nyi random graphs
$G(N, 1/2)$. We emphasize that we are working with the $\mathbb{F}_2$ 
rather than real or complex numbers.

\subsubsection{ The rank of $\Gamma$ is even}

An interesting numerical observation made in Ref.~\cite{Xu2025prb}
is that in each state trajectory of unitary circuit, 
$g_3$ always changes by multiples of two. From the perspective of
graph theory, it simply follows from the fact that the 
rank of any adjacency matrix over $\mathbb{F}_2$ is always even. 
This is a well-know result about skew-symmetric matrix. 
We briefly recall the constructive proof below. 

Since we work over $\mathbb{F}_2$, obviously we have 
$\vec{x}^{\mathsf T}\Gamma \vec{x}=0$ for all $\vec{x}$.
Suppose \(\Gamma\) is nonzero. Pick any nonzero vector $\vec{v}$ such that
$\vec{v}$ is not in the kernel of $\Gamma$. Then there 
exists a vector $\vec{v}_*$ satisfying
\begin{equation}
    \vec{v}_*^{\mathsf T}\Gamma\,\vec{v}=1 \pmod 2.
\end{equation}
Restricting \(\Gamma\) to the subspace spanned by \(\{\vec{v}, \vec{v}_*\}\), the bilinear form takes the standard symplectic block form:
\begin{equation}
\Gamma|_{\mathrm{span}\{\vec{v}, \vec{v}_*\}}
=
\begin{pmatrix}
0 & 1 \\
1 & 0
\end{pmatrix},
\end{equation}
whose rank is 2. Now consider any vector \(\vec{w}\) linearly independent of \(\vec{v}\) and \(\vec{v}_*\). We may adjust \(\vec{w}\) so that it becomes orthogonal to both \(\vec{v}\) and \(\vec{v}_*\) without changing its linear independence. Indeed,
if \(\vec{w}^{\mathsf T}\Gamma\,\vec{v}=1\), replace \(\vec{w}\) by \(\vec{w}+\vec{v}_*\). 
This flips the value of the coupling to \(\vec{v}\) while preserving independence.
Similarly, if \(\vec{w}^{\mathsf T}\Gamma\,\vec{v}_*=1\), replace \(\vec{w}\) by \(\vec{w}+\vec{v}\). This eliminates the coupling to \(\vec{v}_*\).

After these substitutions, we have
$\vec{w}^{\mathsf T}\Gamma\,\vec{v}
= \vec{w}^{\mathsf T}\Gamma\,\vec{v}_*
= 0$.
Once $\vec{w}$ is fixed in this manner, the non-degeneracy of the form on the quotient space guarantees the existence of a conjugate vector \(\vec{w}_*\) satisfying 
$\vec{w}_*^{\mathsf T}\Gamma\,\vec{w}=1$,
$\vec{w}_*^{\mathsf T}\Gamma\,\vec{v}=0$,
and $\vec{w}_*^{\mathsf T}\Gamma\,\vec{v}_*=0$.
Thus, the restriction of $\Gamma$ to $\mathrm{span}\{\vec{v},\vec{v}_*,\vec{w},\vec{w}_*\}$ decomposes into two 
identical $2\times 2$ blocks, each of rank 2.
Repeating this procedure iteratively exhausts the entire vector space. The 
matrix \(\Gamma\) is therefore brought into a block-diagonal form consisting 
of \(2\times 2\) blocks \(\begin{pmatrix}0&1\\1&0\end{pmatrix}\). Hence, the 
rank of \(\Gamma\) is twice the number of such blocks, and is therefore an 
even integer. This block-diagonal decomposition is precisely the standard form 
of a symplectic basis, where the vectors are grouped into conjugate pairs 
$\{(\vec{v},\vec{v}_*),(\vec{w},\vec{w}_*),\dots\}$.

\subsubsection{The Even--Odd Effect}

\begin{figure}[bt]
    \centering
    \includegraphics[width=1\linewidth]{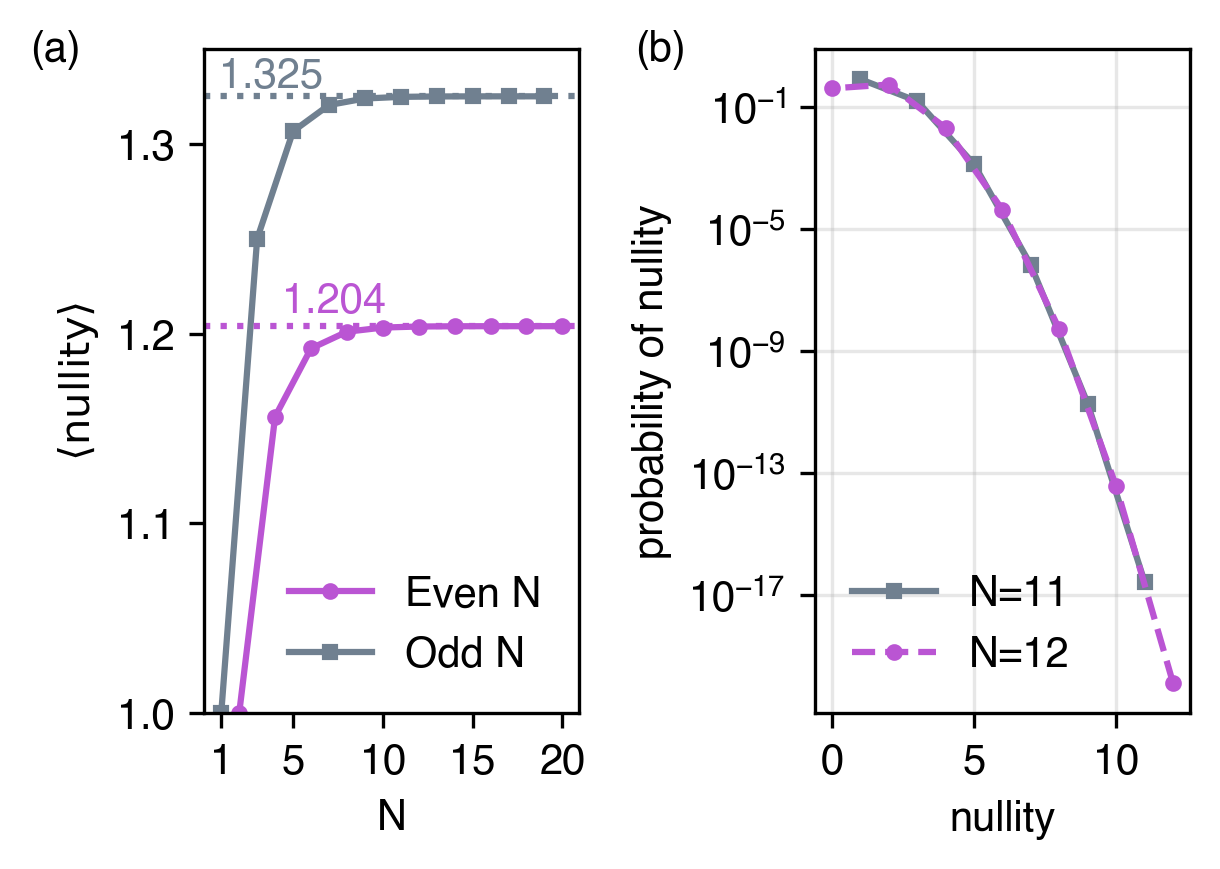}
    \caption{(a) The dependence of the 
    expectation value of $\langle \dim\ker \Gamma \rangle$ on qubit number $N$. 
    Already at $N=10$, the result coincides well with the asymptotic limit (dotted lines).
    (b) The distribution of nullity for $N=11$ and $12$.}
    \label{fig_g3}
\end{figure}

Here we count the number of rank-$2m$ adjacency matrix with $m$ an integer. 
Since the adjacency matrix can be
transformed into blocks of $\begin{pmatrix}
     0 & 1 \\
     1 & 0
\end{pmatrix}$, a new graph can be obtained by 
congruence, i.e., $h\Gamma^{2m}h^{\mathrm{T}}$, from a standard graph $\Gamma^{2m}$.
Therein, $h$ belongs to the general linear
matrix group over $\mathbb{F}_2$, $h\in \mathrm{GL}(N,\mathbb{F}_2)$,
and $\Gamma^{2m}$ is the block-diagonal matrix
\begin{equation}
\Gamma^{2m} = \begin{bmatrix}
 J_{2m} & 0\\
 0 & 0_{N-2m}
\end{bmatrix},
\end{equation}
where $J_{2m}$ is the symplectic matrix over $\mathbb{F}_2$ 
\begin{equation}
J_{2m}=\begin{bmatrix}
    0_m & \mathbb{I}_m \\
    \mathbb{I}_m & 0_m
\end{bmatrix}.
\end{equation}
To obtain the number of rank-$2m$ graphs, we need to know
the subgroup of $\mathrm{GL}(N,\mathbb{F}_2)$ keeping 
the standard form $\Gamma^{2m}$ invariant under congruence, i.e., the subgroup
of $h$ satisfying $h\Gamma^{2m}h^{\mathrm{T}}=\Gamma^{2m}$.

Write such matrix $h$ in the form of
$\begin{bmatrix}
     A_{2m} & B \\
     C & D_{(N-2m)}
\end{bmatrix}$.
The condition $h\Gamma^{2m}h^{\mathrm{T}}=\Gamma^{2m}$ implies:
\begin{enumerate}
    \item $A J_{2m} A^{\text{T}}=J_{2m}$ so that $A$ can be an
    arbitrary element of the $2m$-dimensional symplectic group over
    $\mathbb{F}_2$, $\mathrm{Sp}(2m,\mathbb{F}_2)$.
    \item $A J_{2m} C^{\text{T}}=0$, which forces $C=0$.
    \item $B$ can be an arbitrary $2m\times (N-2m)$ matrix. 
    \item $D$ can be an arbitrary matrix as long as
    $h$ is invertible. Thus, it can be any one of $\mathrm{GL}(N-2m, \mathbb{F}_2)$.
\end{enumerate}
Therefore, the size of the subgroup of 
$\mathrm{GL}(N, \mathbb{F}_2)$ leaving $\Gamma^{2m}$ invariant equals the 
product of the number of each component. The size of symplectic matrix is 
\begin{equation}
     \abs{\mathrm{Sp}(2m, \mathbb{F}_2)} = 2^{m^2}\prod_{i=1}^{m}(2^{2i}-1).
\end{equation}
The size of the $\mathrm{GL}(N, \mathbb{F}_2)$ is 
\begin{equation}
\abs{\mathrm{GL}(N,\mathbb{F}_2)} = 2^{N(N-1)/2}\prod_{i=1}^{N}(2^i-1).   
\end{equation}
The number of $B$ is simply $2^{2m(N-2m)}$. In the end, the 
number of rank--$2m$ graph is 
\begin{equation}
\label{eq:N2m}
\begin{split}
    N_{2m} & =\frac{2^{-2m(N-2m)}\abs{\mathrm{GL}(N,\mathbb{F}_2)}}
    { \abs{\mathrm{GL}(N-2m,\mathbb{F}_2)}\abs{\mathrm{Sp}(2m, \mathbb{F}_2)}} \\
& = 2^{m(m-1)}\frac{\prod_{i=0}^{2m-1} (2^{N-i}-1)}{\prod_{i=1}^m (4^i-1)}.
\end{split}
\end{equation}

The probability to have a graph with rank $2m$ is 
\begin{equation}
\Pr(2m)= N_{2m}/2^{N(N-1)/2}.
\label{eq:pr_rank}
\end{equation}
The expected nullity follows directly:
\begin{equation}
\label{eq:odd-even}
\begin{split}
\langle \dim \ker \Gamma\rangle  = & \sum_{m=0}^{\lfloor N/2\rfloor} (N-2m)
\Pr(2m) \\
& \xrightarrow{\text{large}\, N}  \begin{cases}
    1.204, \quad  & N \text{ is even}; \\
    1.325, \quad   & N \text{ is odd}.
\end{cases}
    \end{split}
\end{equation}
The asymptotic limits agree with the odd--even effect 
depicted in Fig.~\ref{fig:system}(b). In Fig.~\ref{fig_g3}(a) we
show that the expected nullity 
converges to the asymptotic limit remarkably fast. 
In Fig.~\ref{fig_g3}(b) we plot the probability distribution of
the nullity for two values of $N$. 
It shows that graphs with high nullity are extremely rare.

\section{Graphs for Monitored Random Clifford Circuits}
\label{sec:AED}

Having seen the graph-based approach can solve problems intractable 
for existing methods, here we ask what the graphs look like when measurements
are present. In particular, we want to know how
\emph{locality}, character of the brickwork layout,
which is hidden in the output of deep unitary circuits,
is retrieved partially or completely by measurements
in the volume- or area-law phases, respectively.
We shall see that the ensemble average of
adjacency matrix elements $\langle \Gamma_{i,j}\rangle$ 
offers a transparent 
view of this process, yet still not enough to narrow down the
output-state distribution.

\begin{figure}[tb]
    \centering
    \includegraphics[width=\linewidth]{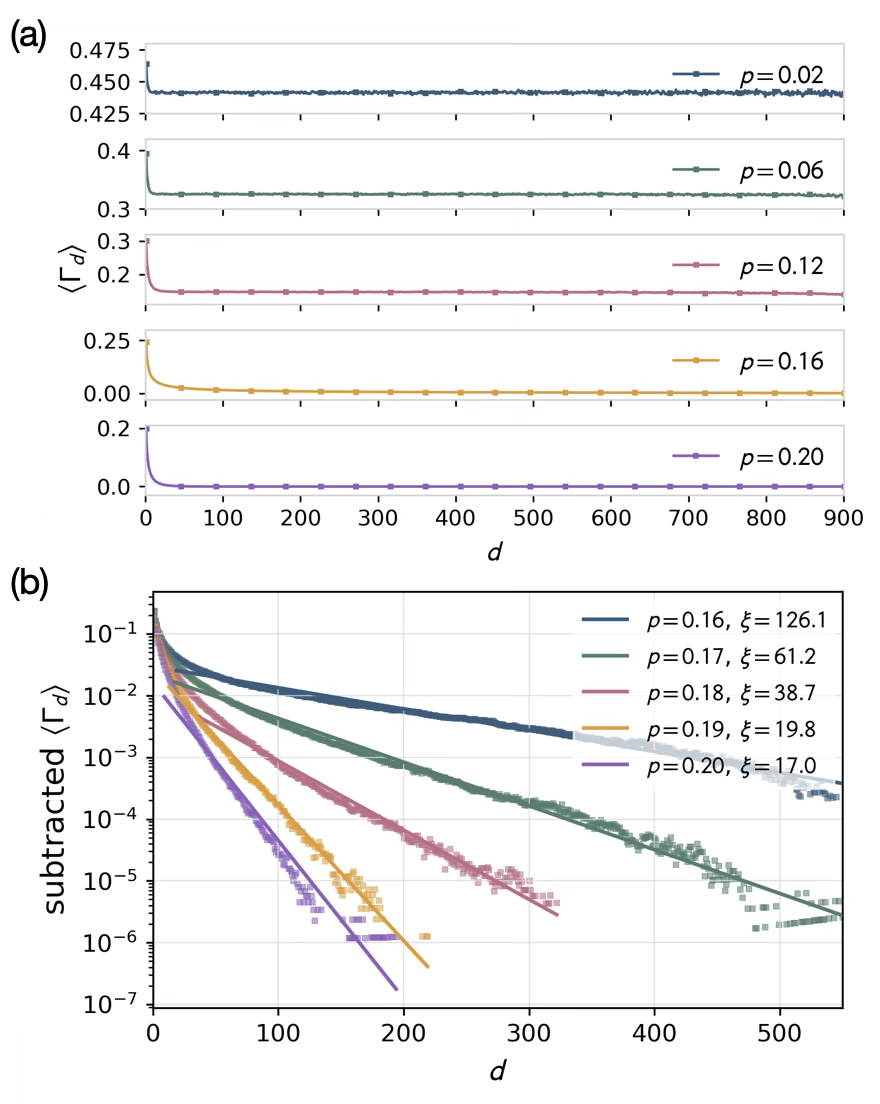}
    \caption{(a) Averaged edge density (AED) for $N=1000$ qubits and values of $p$ detailed in the plots.
    Graphs of the volume-law phase ($p=0.02, 0.06,0.12$) have a homogeneous 
    long-range AED $\langle \Gamma_d\rangle>0$, which drops to 
    zero near the critical point ($p=0.16$) and in the area-law
    phase ($p=0.20$). (b) The decay of AED $\langle \Gamma_d\rangle$ subtracted by
    the long-range AED (tiny but nonzero due to finite size) in the area-law phase. 
    The data fits the exponential decay $\exp(-d/\xi)$.}
    \label{fig:AED_homo}
\end{figure}

\subsection{Long- and Short-range Behavior}
\label{subsec:homoAED}

\begin{figure}[bt]
    \centering
    \includegraphics[width=0.85\linewidth]{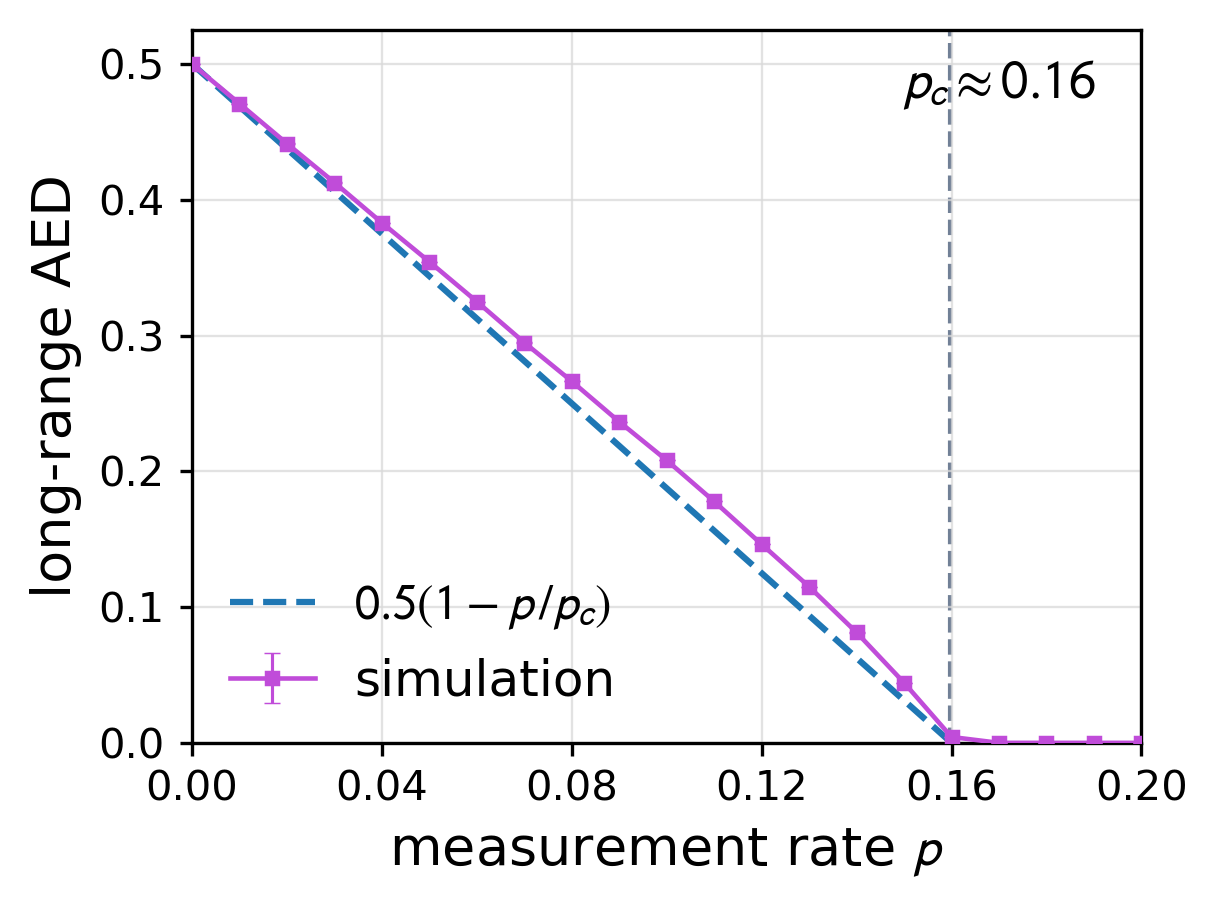}
    \caption{The homogeneous long-range AED (solid curve) decays as a function of $p$. It is approximated by a straight dashed curve $0.5(1-p/p_c)$ with
    $p_c\approx 0.16$. }
    \label{fig:AED_p}
\end{figure}

As explained in Fig.~\ref{fig:system}(c), quasi-translation 
invariance in the bulk of the 1D chain implies that elements
$\langle\Gamma_{i,j}\rangle$ with fixed distance $d=\abs{i-j}$ behaves identically. 
We thus consider the averaged edge density (AED)
\begin{equation}
    \langle\Gamma_d\rangle=\frac{1}{\sum_{i<j}\delta_{d, i-j}}\sum_{i<j}\langle \Gamma_{i,j}\rangle\delta_{d, i-j}.
\end{equation}
We plot AED as a function of $d$ for various
measurement rate $p$ in Fig.~\ref{fig:AED_homo}(a).
The figure clearly demonstrates a \emph{homogeneous} finite
long-range AED in the volume-law phase.
That is, the AED curves are flat against $d$ except for the
region where $d$ is small. Values of the homogeneous long-range AED 
decreases with $p$ and drops to
zero when the system enters the area-law phase, 
as shown in Fig.~\ref{fig:AED_p}. By comparing $\langle\Gamma_d(p)\rangle$
with $0.5(1-p/p_c)$, the figure also indicates that
the decay of long-range AED is almost linear.

The signature of locality is reflected in the decaying short-range AED. 
In the area-law phase,  the decay becomes exponential in $d$, 
as illustrated in Fig.~\ref{fig:AED_homo}(b) for several values of
$p$, with fits to the form
$\exp(-d/\xi)$.
In the volume-law phase, AED decays with $d$ before saturating
to the long-range plateau, as shown in Fig.~\ref{fig:AED_homo}(a). 
But this decay is not simply exponential decay owing to a backflow
from the long-range part.

To summarize, when measurements are introduced, the graphs
of output states in the volume-law phase display a 
clean feature: A homogeneous long-range AED, which signals
the non-locality of entanglement structure.
Meanwhile, locality is manifested in the decay of  
short-range AED with $d$, and in the suppression of the 
homogeneous long-range AED as $p$ increases. 
In this sense, the long-range AED 
can be viewed as the order parameter for the volume-law phase.

\subsection{What GHZ Statistics Hides}
\label{subsec:ghz_ambiguity}

The above numerical results raise several new questions. First, can the 
graph representation shed light on new understandings about the MIPT? 
Second, how does the homogeneous long-range AED, a common feature of 
the volume-law phase, connect to the GHZ plateau? 
And, ultimately, what general features can we infer about individual output states? 
The presence of such a homogeneous long-range structure 
points toward a simple and universal answer.

A natural hypothesis is that the output-state graph contains a subgraph
of uniform edge density denoted by $r$, specifically, an Erd\H{o}s--R\'{e}nyi random
graphs $G(N_{\sub}, r)$, where $N_{\sub}<N$ denotes 
the size the subgraph. If this holds, then, $r\binom{N_{\sub}}{2}/\binom{N}{2}\approx r N_{\sub}^2/N^2$ may match the
numerically observed long-range AED. 
The reason it is a proper subgraph is
simply that vertices isolated by measurements contribute no edges.
At the same time, this structure must be consistent with the
GHZ plateau depicted in Fig.~\ref{fig:system}(b). 
Unfortunately, as we demonstrate in App.~\ref{app:Gnp},
graph states associated with $G(N,r)$ for any $r< 1/2$ shares the same GHZ statistics 
as those of $G(N,1/2)$ in the large-$N$ limit. 
This ambiguity necessitates a more detailed examination of the 
circuit dynamics directly at the level of graph representations.

\section{Graph Operations for Gates and Measurements}
\label{sec:graph_operation}

Here we introduce the graph-update rules 
corresponding to two-qubit Clifford gates and measurements. 
What is presented in this section is not new but necessary for the
discussions of monitored circuits.

\subsection{An Equivalent Circuit}

We begin from an overview of the Clifford group. The single-qubit
Clifford group $\mathcal{C}_1$ has 24 elements (up to global phase)
and is generated by the Hadamard gate $H$ and the phase gate $S$. The two-qubit 
Clifford group $\mathcal{C}_2$ has 11520 elements (up to global phase). It is
generated by $H$, $S$ and the two-qubit $\mathrm{CZ}$~\eqref{eq:cz}.
Obviously, $\mathcal{C}_2$ has a trivial
subgroup, $\mathcal{C}_1\otimes \mathcal{C}_1$, which 
has $24^2=576$ elements. We consider the orbit space of $\mathcal{C}_2$ 
under local Clifford conjugation, i.e., for any $g\in \mathcal{C}_2$ and $h\in\mathcal{C}_1\otimes \mathcal{C}_1$ we identify 
$g\sim h gh^{\dagger}$. Since $\mathrm{CZ}$ is the unique 
two-qubit generator, we calculate 
$\left(U_1\otimes U_2\right)^\dagger \mathrm{CZ} \left(U_1\otimes U_2\right)$
and obtain 9 possible results 
\begin{equation}
    \mathrm{G}^{P,Q}=\frac{1}{2}\left(
    \mathbb{I}\otimes \mathbb{I}+P\otimes \mathbb{I}
    +\mathbb{I}\otimes Q
    -P\otimes Q
\right).
\end{equation}
where $P,Q\in\{X,Y,Z\}$. In the end, there are 20 cosets of
$\mathcal{C}_1\otimes\mathcal{C}_1$ in $\mathcal{C}_2$ 
classified into 4 families:
$\mathbb{I}$, $\mathrm{SWAP}$, $\mathrm{CZ}$, and $\mathrm{SWAP}\cdot\mathrm{CZ}$, and each of them
covers 1, 1, 9, 9 cosets, respectively. For example,
a gate in the $\mathrm{CZ}$ class can be expressed as 
$(V_1\otimes V_2)G^{P,Q}$. 

Note that $(V_1\otimes V_2)$ can be absorbed into the
next layer of gates (or measurements). Repeating this
logic, every random single-qubit Clifford gate is either 
pushed forward to the last circuit layer, 
or absorbed by a $Z$-measurement (which then may
become $X$- or $Y$-measurement). What
left in the bulk of the circuits are one of
ten gates $\{\mathbb{I}_4\}\cup\{G^{P,Q}\}_{(P,Q)}$,
followed by a $\mathrm{SWAP}$ with probability $1/2$.
To summarize, the brickwork circuit is equivalent to
\begin{description}
    \item[Gates] Either do nothing or implement one of the nine distinct gates $G^{P,Q}$, each with probability $1/10$. Then, with probability $1/2$, apply a SWAP gate.
    \item[Measurements] With probability $p/3$, perform a measurement in one of the bases $\{X,Y,Z\}$.
\end{description}
We remark that at the last circuit layer, 
the single-qubit Clifford gates are not guaranteed to follow the gauge of canonical extended
graph state. Thus, a reshuffle of graph is necessary within the 
formalism of canonical extended graph state. But this step does not
change anything about quantum entanglement.

\subsection{Graph Updating Rules}
\label{subsec:table}

Following the notations of Ref.~\cite{Hu:2022aa}, we apply
$G^{P,Q}$ on an ordered pair of qubits $(v_1, v_2)$,
so that $P$ and $Q$ are referred to $v_1$ and $v_2$, respectively.
The problem of casting state $G^{P,Q}\ket{\Gamma}$ into 
canonical extended graph state was 
already addressed in Refs.~\cite{Elliott:2008aa,Hu:2022aa}. 
A brief summary of how these rules are derived are given in 
App.~\ref{app:measurements}.
The results are reproduced 
in Tab.~\ref{tab:1}, where we keep only graph operations
since the additional local Clifford gates can be pushed forward
to the final circuit level. 

\begin{table}[t]
\centering
\begin{tabular}{c  @{\hspace{1em}} c}
\hline\hline
Gates $(P, Q)$ & graph operation on $\Gamma$ \\
\hline
(Z,Z) & $T_{1,2}$  \\
(Z,X) & $\prod_{i\in \mu(2)}T_{1,i}$ \\
(Z,Y) & $\prod_{i\in \mu^{*}(2)} T_{1,i}$ \\
(X,X): (1$\sim$2) & $T_{1,2}\prod_{i\in \mu^{*}(1)}\prod_{j\in\mu^{*}(2)} T_{i,j}$ \\
(X,X): (1$\nsim$2)  & $\prod_{i\in \mu(1)}\prod_{j\in\mu(2)} T_{i,j}$ \\
(Y,X): (1$\sim$2) & $\prod_{j\in\mu^{*}(1)\triangle\mu^{*}(2)}T_{1,j}\prod_{(i,j)\in \mu(1)}T_{i,j}$ \\
(Y,X): (1$\nsim$2) & $\prod_{(i,j)\in\mu^{*}(1)\triangle\mu(2)}T_{i,j}\prod_{(i,j)\in \mu^{*}(1)}T_{i,j}$ \\
(Y,Y): (1$\sim$2) & $\prod_{(i,j)\in\mu^{*}(1)}T_{i,j}\prod_{(i,j)\in \mu^{*}(2)}T_{i,j}$ \\
(Y,Y): (1$\nsim$2) & $\prod_{j\in\mu^{*}(2)}T_{1,j}\prod_{(i,j)\in \mu(1)}T_{i,j}$ \\
\noalign{\vskip 4pt}
\hline\hline
Measurements &  \\
\hline
Z & isolate $v_1$ \\
Y & $\prod_{(i,j)\in\mu(1)}T_{i,j}$, isolate $v_1$ \\
X & $\prod_{i\in\mu^*(1)}\prod_{j\in\mu^*(2)}T_{i,j}$, isolate $v_1$\\
\noalign{\vskip 3pt}
\hline\hline
\end{tabular}
\caption{(Upper) Rules for obtaining the graph of $G^{P,Q} \ket{\Gamma}$
from $\Gamma$. Therein, 1$\sim(\nsim)$2 denotes the case where
$v_1$ and $v_2$ are connected (dis-connected) in $\Gamma$,
$T_{i,j}$ denotes the map $\Gamma_{i,j}\rightarrow\Gamma_{i,j}\oplus 1$, 
$\mu^*(1)\equiv \{v_1\}\cup \mu(1)$,
$\triangle$ denotes the symmetric 
difference of two sets, $(i,j)\in\mu^*(1)$ means 
pairs of vertices in $\mu^*(1)$, etc. 
(Lower) Rules for post-measurement states conditioned on obtaining
outcome +1 in the measurement of Pauli operators on vertex $v_1$. 
Therein, ``isolated $v_1$'' means 
deleting all edges incident to $v_1$. For the case of $X$--measurement 
one needs to specify an arbitrary $v_2 \in \mu(1)$.}
\label{tab:1}
\end{table}


Graph operations corresponding to measurements on qubit $v_1$ 
are also included 
in Tab~\ref{tab:1}. Therein, all measurement outcomes are assumed as
+1, since the post-measurement state of outcome -1 differs 
by only products of single-qubit Clifford gates.
These operations are derived firstly in Ref.~\cite{Hein:2004aa,Anders:2006aa}.
In particular, $X$- and $Y$-measurements may induce
entanglement among other qubits, see the edge toggles 
shown in Tab~\ref{tab:1}.

At last, we note that in the case of $(Y,Y):(1\nsim 2)$, the 
graph operation displayed in Tab.~\ref{tab:1} is not 
symmetric w.r.t. $(v_1, v_2)$
although the gate is. Asymmetry also appears for the
$X$-measurement, where one needs to specify one neighbor of
the vertex being measured. This asymmetry roots in the
local-Clifford-equivalence among different graph states.

\section{Revisit the Volume-Law Phase and MIPT}

Our study of monitored random Clifford circuits 
is built upon a key observation: There are three distinct types 
of edge toggles, exemplified in Fig.~\ref{fig:toggles}. 
Analyzing the effect of each leads us to a completely new 
perspective on both the volume-law phase 
and MIPT.

\begin{table}[b]
\centering
\begin{tabular}{c  @{\hspace{1em}} c}
\hline\hline
Gates $(P, Q)$ & adjacency matrix: $\Gamma_{i,j}\rightarrow \Gamma_{i,j}+(\cdots)$ \\
\hline
(Z,Z) & $\delta_{i,1}\delta_{j,2}+\pi_{ij}$  \\
(Z,X) & $\delta_{i,1}\Gamma_{j,2}+\pi_{ij}$ \\
(Z,Y) & $\delta_{i,1}(\delta_{j,2}+\Gamma_{j,2})+\pi_{ij}$ \\
(X,X): (1$\sim$2) & $(\Gamma_{i,1}\Gamma_{j,2}+\delta_{i,1}\Gamma_{j,2}+\delta_{j,2}\Gamma_{i,1})+\pi_{ij}$ \\
(X,X): (1$\nsim$2)  & $\Gamma_{i,1}\Gamma_{j,2}+\pi_{ij}$ \\
(Y,X): (1$\sim$2) & $\Gamma_{i,1}\Gamma_{j,1}+\delta_{i,1}(\Gamma_{j,1}+\Gamma_{j,2}+\delta_{j,2})+\pi_{ij}$ \\
(Y,X): (1$\nsim$2) & $\Gamma_{i,2}\Gamma_{j,2}+\Gamma_{i,2}(\delta_{j,1}+\Gamma_{j,1})+\pi_{ij}$ \\
(Y,Y): (1$\sim$2) & $(\Gamma_{i,1}+\delta_{i,1})(\Gamma_{j,1}+\delta_{j,1})+
\pi_{12}$ \\
(Y,Y): (1$\nsim$2) & $\Gamma_{i,1}\Gamma_{j,1}+\delta_{i,1}(\Gamma_{j,2}+\delta_{j,2})
+\pi_{ij}$ \\
\noalign{\vskip 4pt}
\hline\hline
Measurements &  \\
\hline
Z & isolate $v_1$ \\
Y & $\Gamma_{i,1}\Gamma_{j,1}$, isolate $v_1$\\
X & $(\Gamma_{i,1}+\delta_{i,1})(\Gamma_{j,2}+\delta_{j,2})+\pi_{ij}$,
isolate $v_1$\\
\noalign{\vskip 3pt}
\hline\hline
\end{tabular}
\caption{Elements of the adjacency matrix will be updated to 
$\Gamma_{ij}+\left(\cdots\right)$, where terms in the bracket is given
in the table. All additions are modulo two. $\pi_{ij}$ and $\pi_{12}$ means permutation 
$(i\rightleftharpoons j)$ and $(1\rightleftharpoons 2)$ of
the preceding \emph{one} term, respectively.}
\label{tab:2}
\end{table}

\subsection{Three Kinds of Edge Toggles}
\label{subsec:toggles}

Table~\ref{tab:1} reveals that local gates can 
induce highly nonlocal graph updates. Let us assume 
$\Gamma$ is sparse so that long-range 
AED (denoted by $r$) can be viewed as a perturbative index. In this case, an edge toggle ($\Gamma_{i,j}:0\rightleftharpoons 1$) 
is more likely to create a new edge ($0\rightarrow 1$)
than deleting one.
Then, if an edge toggle is conditioned on $\ell$ preexisting \emph{long-range} 
edges, 
the probability for it to established a new edge is 
at the order of $O(r^\ell)$.  
Based on this, we identify three kinds of edge toggles 
labeled by $C_\ell$ with
$\ell\in\{0,1,2\}$.

An alternative way to manifest this classification is to
translate Tab.~\ref{tab:1} into rules for the adjacency matrix. 
We present the results in
Tab.~\ref{tab:2} where one can see terms of different degrees in $\Gamma_{i,1(2)}$ or $\Gamma_{j,1(2)}$
for $v_{i(j)}$ that can be far from $v_{1(2)}$.
$C_1$ and $C_2$ toggles thus correspond to the first- and second-degree terms. 
($C_0$ toggles cannot be naively ascribed to the zeroth-degree terms.)
In addition, one may establish a complete description of
the graph evolution basing on Tab.~\ref{tab:2}. 
But this is not in the scope of this Article.

\begin{figure}
    \centering
    \includegraphics[width=\linewidth]{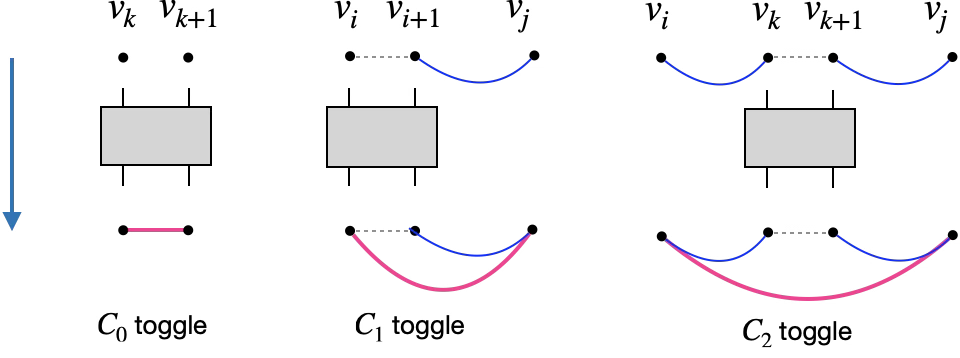}
    \caption{Schematic illustrating the three types of edge toggles and the typical 
    configurations through which a new edge (red) is established. Time evolves downward. 
    Dashed lines represent edges whose existence is conditional.}
    \label{fig:toggles}
\end{figure}

In Fig.~\ref{fig:toggles} we illustrate three
typical configurations, one for each $C_{\ell}$. In general, they are introduced below.
\begin{description} 
\item[$C_0$] Nearest-neighbor connection $\Gamma_{i,i+1}$ 
will be toggled by gates on
$(v_i, v_{i+1})$ if $(P,Q)$ is one of $(Z,Z)$, $(Z,Y)$, $(Y,Z)$, and 
either $(X,X)$ if $\Gamma_{i,i+1}=1$ or $(Y,Y)$ otherwise.
The effect of $C_0$ toggles will be discussed in Sec.~\ref{subsec:C0}.

\item[$C_1$]  A gate on $(v_i, v_{i\pm 1})$ may connect a 
remote pair ($v_i$, $v_j$) provided that
$v_j\in\mu(i\pm 1)$. Similarly,  a gate on $(v_j, v_{j\pm 1})$ also works
provided that $v_{i}\in\mu(j\pm 1)$. The
gate $G^{P,Q}$ should be one of the following seven: 
$(Z,X)$, $(X,Z)$, $(Z,Y)$, $(Y,Z)$, $(Y,X)$, 
$(X,Y)$, and either $(X,X)$ if the gate-acting
nearest-neighbor pair is connected already
or $(Y,Y)$ otherwise. In the case 
of $(1\sim 2)$, a gate may also delete existing edges. 
For example, $(Y,Y)$ employs the operation 
$\prod_{(i,j)\in \mu^*(2)}T_{i,j}$, which strips $v_2$ of all its neighbors
and reassigns them to $v_1$. 
The effect of $C_1$ toggles will be discussed in Sec.~\ref{subsec:pc}.

\item[$C_2$] 
A remote pair ($v_i$, $v_j$) can also be connected through
a gate on neither of them. This is the case where a gate $G^{P,Q}$ 
on $(v_k, v_{k+1})$ for any $k$ is one from the following four:
$(X,X)$, $(X,Y)$, $(Y,X)$, $(Y,Y)$, and simultaneously 
$v_i\in \mu(k)$ and $v_j\in\mu(k+1)$ (the configuration shown in Fig.~\ref{fig:toggles}).
Other possible configurations is when both $v_i$ and $v_j$ are
in $\mu(k)$ or $\mu(k+1)$, especially in $X$- and $Y$-measurements. 
We discuss $C_2$ toggles in Sec.~\ref{subsec:C2toggle}.
\end{description}

\subsection{$C_0$ Toggles and the Nearest-Neighbor Connections} 
\label{subsec:C0}

\begin{figure}[tb]
    \centering
    \includegraphics[width=0.9\linewidth]{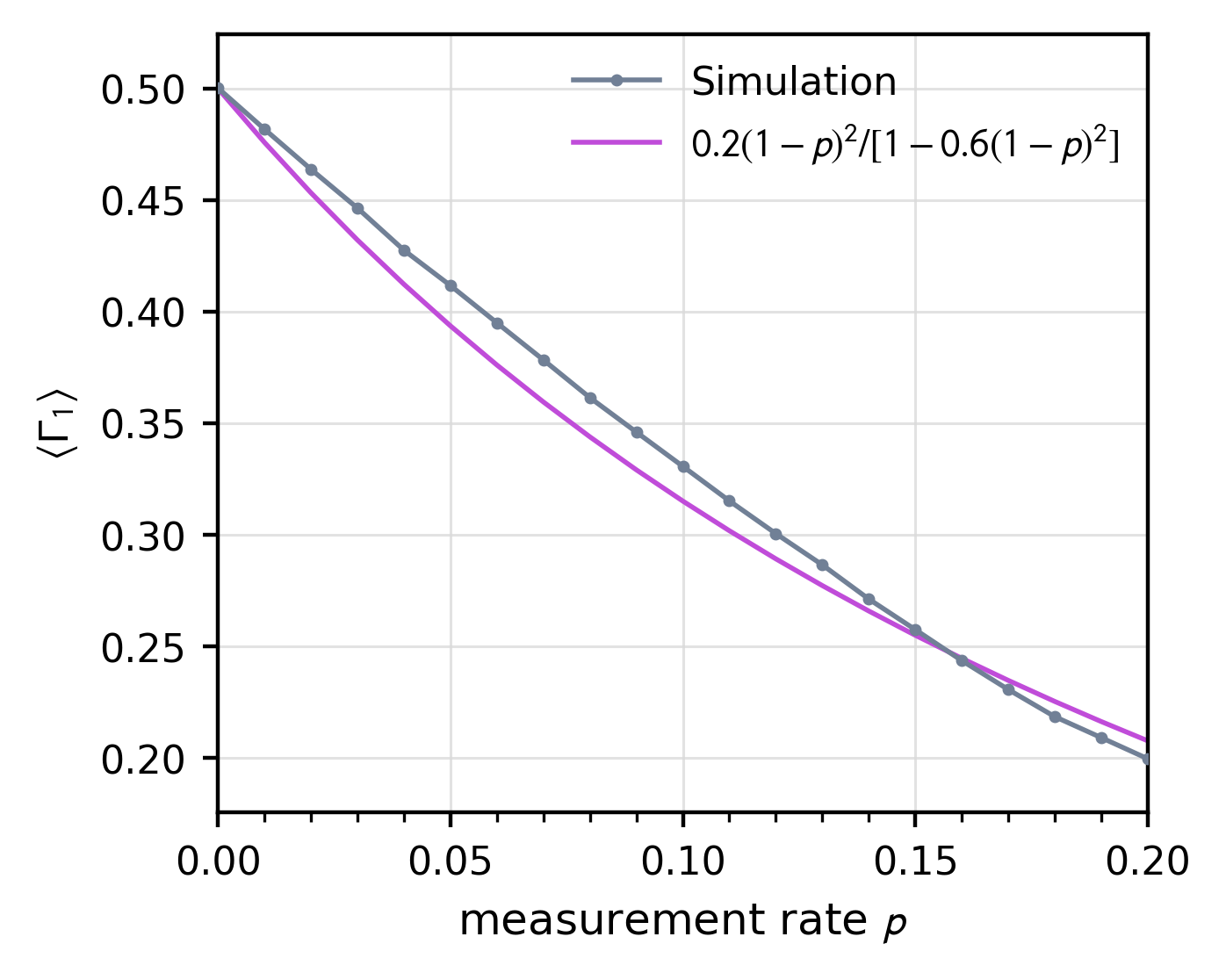}
    \caption{Comparison between numerical simulations and the
    mean-field estimate Eq.~\eqref{eq:r1} for $\langle \Gamma_1\rangle$.
    The two curves cross in the vicinity of 
    the critical point $p_c\approx 0.16$.}
    \label{fig:gamma1}
\end{figure}

$C_0$ toggles act on the nearest-neighbor 
edges. We assume an edge density $\langle \Gamma_{i,i+1}\rangle=\langle\Gamma_1\rangle$, 
$\forall i$, and consider a mean-field determination of it. 
For now we ignore the other two kinds of toggles.
As summarized in Sec.~\ref{subsec:toggles}, 
there are four $G^{P,Q}$ that contribute to $C_1$ toggles. 
Thus, the probability that a gate
on $(v_i, v_{i+1})$ toggles $\Gamma_{i,i+1}$ is $0.4$.
Since this gate occurs every two levels of the brickwork, 
we set the probability by $0.2$ on average. 
Therefore, $\langle \Gamma_1\rangle$ is updated by one level of circuit to
\begin{equation}
\langle \Gamma'_1\rangle=\langle \Gamma_1\rangle-0.2 
\langle \Gamma_1\rangle+0.2(1-\langle \Gamma_1\rangle).
\end{equation}
Then, the random measurements 
keep only $(1-p)$ vertices intact so that the equilibrium condition 
can be formulated as $\langle \Gamma'_1\rangle(1-p)^2=\langle \Gamma_1\rangle$, which leads to 
\begin{equation}
\label{eq:r1}
    \langle \Gamma_1\rangle=\frac{0.2(1-p)^2}{1-0.6(1-p)^2}.
\end{equation} 
In Fig.~\ref{fig:gamma1}, we compare
this mean-field estimate with the result of numerical simulations
on $N=1000$ qubits.
The figure shows good overall agreement, although
Eq.~\eqref{eq:r1} underestimates $\langle \Gamma_1\rangle$ in the volume-law phase and
overestimates it in the area-law phase. Notably, 
the crossing point occurs at $p\approx0.16$, precisely where the critical point of
MIPT is expected. This observation strongly suggests
that the graph-based mean-field theory could be used to
locate the critical point. We shall return to it in Sec.~\ref{subsec:pc}.

\subsection{$C_2$ Toggles Underlie the Volume-Law Phase}
\label{subsec:C2toggle}

$C_2$ toggles are distinguished by nonlocality. 
We assume a sparse graph
characterized by a homogeneous 
long-range AED $r$,
and consider the probability for
a disconnected remote pair $(v_i\nsim v_j)$ to be connected through
$C_2$ toggles generated in one circuit level.
As summarized in Sec.~\ref{subsec:toggles}, 
a $C_2$ edge toggle may come from 
four kinds of $G^{P,Q}$ .
There are $N/2$ independent gates in each circuit layer, 
and every gate has a chance of 
$q=0.4 r^2$ to establish a $C_2$ toggle for $(v_i\nsim v_j)$.
Thus, they are connected eventually, $\Gamma_{ij}\rightarrow 1$, 
only if such
$C_2$ toggle occurs for an odd number of gates among all
$N/2$ gates.
The probability for this to happen turns out to be 
\begin{equation}
\label{eq:C2}
\begin{split}
\mathrm{Pr}(\Gamma_{ij}\vert C_2) = & \frac{1}{2}\left[1-(1-0.8r^2)^{N/2}\right] \\[0.8em]
\xrightarrow{\text{large}\, N} & 
\begin{cases}
     0.2 Nr^2\rightarrow 0, & r= o(N^{-1/2}); \\[0.6em]
     1/2,  & r =\omega(N^{-1/2});
\end{cases}
\end{split}
\end{equation}
where $o(N^{-1/2})$ and $\omega(N^{-1/2})$ means 
$r$ grows strictly slower or faster than $N^{-1/2}$, respectively.

\subsubsection{Emergent Erd\H{o}s--R\'{e}nyi
Subgraph $G(N_{\mathrm{sub}}, 1/2)$ }

Equation~\eqref{eq:C2} implies that when AED 
exceeds the scaling $N^{-1/2}$, 
the $C_2$ toggles will immediately lift it to 
$1/2$. The assumption of homogeneous AED can be relaxed to
a subsystem with size $N_\sub=O(N)$, which then leads to 
Erd\H{o}s--R\'{e}nyi
random subgraph $G(N_{\mathrm{sub}}, 1/2)$.
To be compatible with the linear
approximation of long-range AED depicted in Fig.~\ref{fig:AED_p},
we should have $N_{\sub}/N\approx \sqrt{1-p/p_c}$.
Strictly speaking, deviations from
$G(N_{\mathrm{sub}}, 1/2)$ might be found in the 
short-range part. But this part has vanishing weight
in the large-$N$ limit.

\begin{figure}[bt]
\centering
\includegraphics[width=\textwidth]{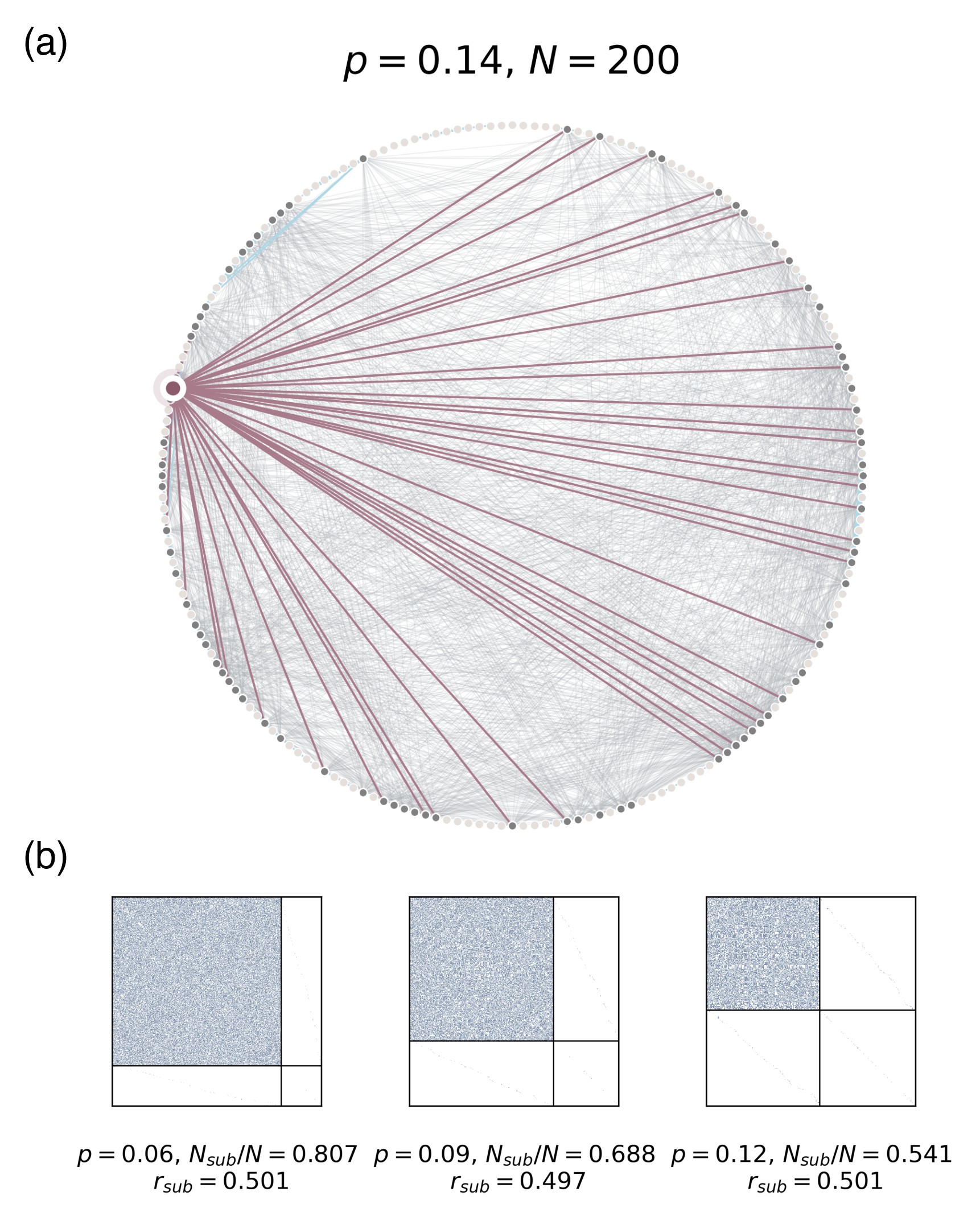}
    \caption{(a) Graph representation of a state produced by a monitored random Clifford 
    circuits with $N=200$ qubits and measurement rate $p=0.14$. 
    Dense-subgraph vertices are indicated by dark markers,
    with one selected vertex and all its incident edges emphasized.
    (b) Adjacency matrices (entries 1 are shown in colored pixel) for 
    three states generated by circuits with 1000 qubits at $p=0.06, 0.09, 0.12$. 
    Size of the subgraph $N_{\sub}$ and the edge density $r_{\sub}$ of the dense 
    subgraph are reported beneath each matrix. Vertices are reordered to 
    put the dense subgraph in the upper left corner.}
    \label{fig:graph_matrix}
\end{figure}

\begin{figure}[bt]
    \centering
    \includegraphics[width=\linewidth]{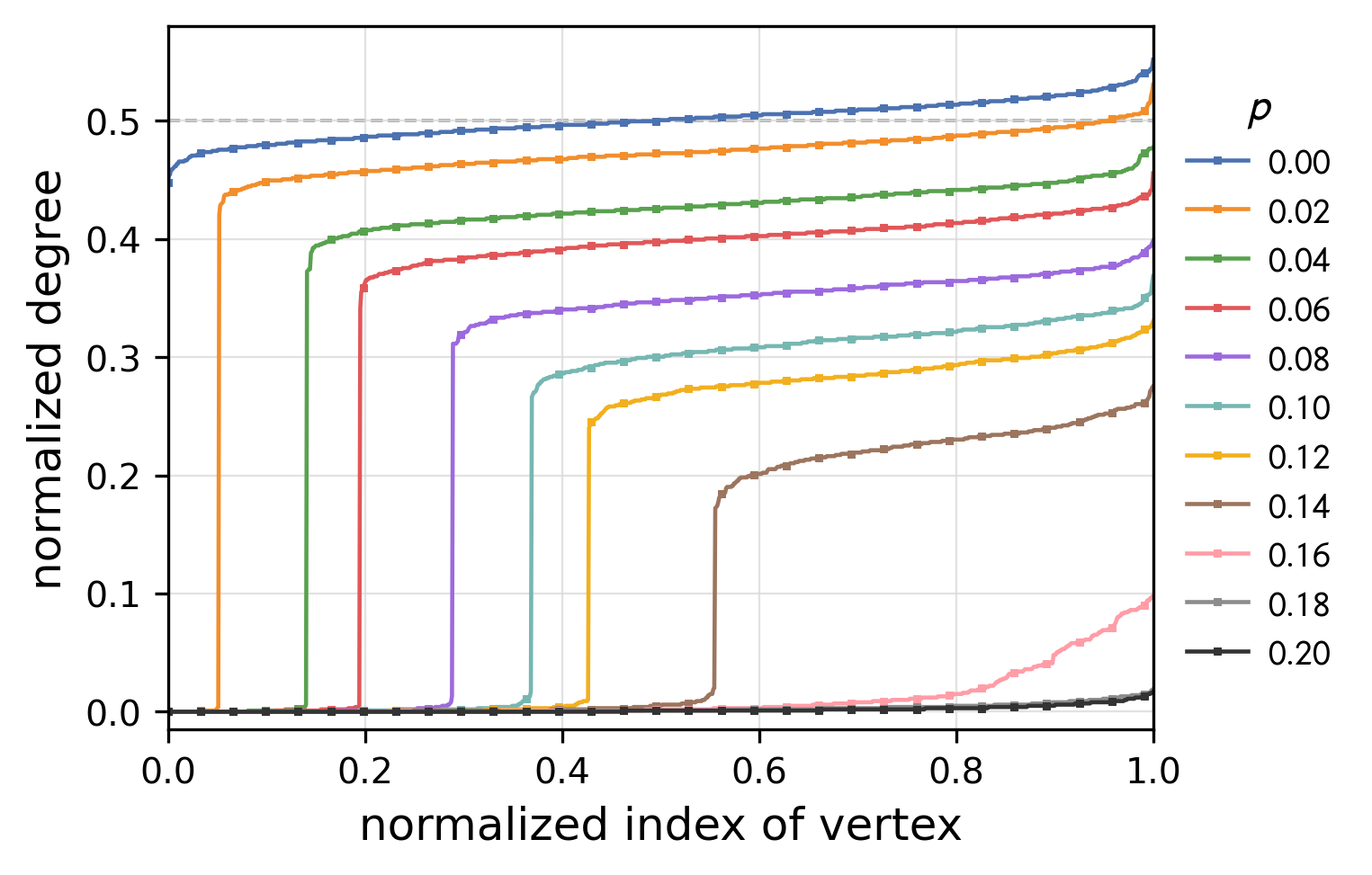}
    \caption{Normalized degree of each vertex as a function of $p$.
    The data is obtained with $N=1000$ and depth $3N$.}
    \label{fig_degree}
\end{figure}

\begin{figure}[b]
    \centering
    \includegraphics[width=0.95\linewidth]{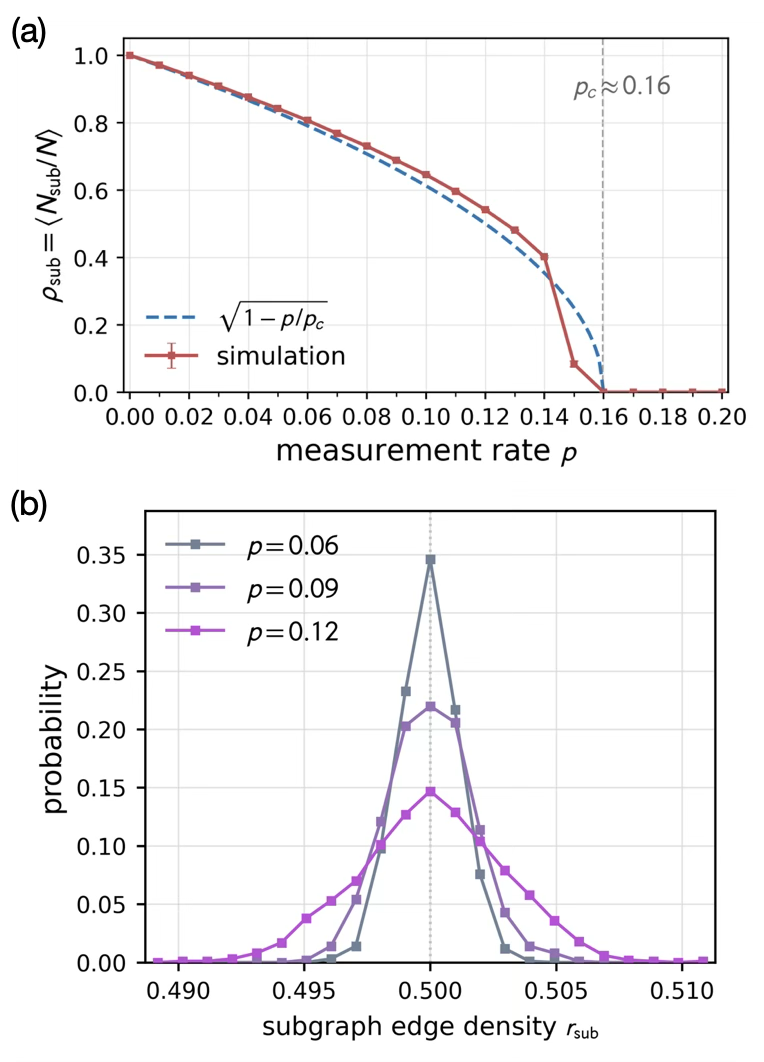}
    \caption{(a) The relative size of the Erd\H{o}s--R\'{e}nyi random
    subgraph $\rho_{\sub}=\langle N_{\sub}/N \rangle$ averaged over 3000 simulations of
    $N=1000$. The data for $p=0.15$ is not reliable because $N_{\sub}$ 
    is so small that the dense subgraph is not recognized faithfully. 
    The dashed curves give the estimate $\rho_{\sub}=\sqrt{1-p/p_c}$, which is 
    derived from the linear approximation of the homogeneous 
    long-range AED in Fig.~\ref{fig:AED_p}. 
    (b) The distribution of normalized degree (number of dense vertices 
    connected to a given vertex divided by $N_{\sub}-1$) of
    each vertex within the dense subgraph. }
    \label{fig:subgraph_size}
\end{figure}

To verify it we plot the graph of an output state of
random circuits with $p=0.14$ and $N=200$ in Fig.~\ref{fig:graph_matrix}(a).
These parameters are chosen to prevent the graph from becoming too dense.
In Fig.~\ref{fig:graph_matrix}(a), the qubit 
chain is arranged in a circle. We mark the 
qubits belong to the dense subgraph by darker colors, 
pick up a vertex and highlight all the edges connecting it to 
other vertices in the subgraph.
In Fig.~\ref{fig:graph_matrix}(b) we choose $N=1000$ and three different
values of $p$, and visualize the adjacency matrices as
pixel plots where colored pixels indicate entries equal to 1.
Therein, numbering of the vertices are reordered to concentrate
the dense subgraph in the upper left corner. These plots
demonstrate a surprising fact that 
all other matrix blocks are extremely sparse [the 
number of entries equal to 1 in these blocks is $O(N)$].
so that the vertices can be classified into ``dense'' and ``sparse'' classes.
This is because, as demonstrated by Eq.~\eqref{eq:C2}, 
the $C_2$ toggles produce a huge gap [between $o(N^{-1/2})$ and $1/2$]
in the scaling of $r$. To verify the classification into dense and sparse 
vertices, we pick up one instance of output state, resort the
vertices by increasing degree (the number of neighbors in the graph),
and plot the normalized degree (divided by $N-1$) 
for various values of $p$ in Fig.~\ref{fig_degree}. In this figure,
we see a sharp cliff in every curve for $p<0.16$, demonstrating
two species of vertices. We also see that the plateau 
of dense vertices is not exactly
flat. This is a finite-size broadening 
further exhibited in Fig.~\ref{fig:subgraph_size}(b).
We therefore confirm that the classification into two species
is valid. This fact inspires a toy model to be presented in the
next subsection.

In Fig.~\ref{fig:subgraph_size}(a), we plot the ratio $N_{\sub}/N$
as a function of $p$, and compare it with the estimate $\sqrt{1-p/p_c}$. 
It shows a good agreement except for
a deviation at $p=0.15$. This is because when $p\approx p_c$ it becomes
difficult for our code to precisely recognize the dense subgraph.
In Fig.~\ref{fig:subgraph_size}(b), we plot the distribution
of the normalized vertex degree (defined within the dense subgraph). 
The plot confirms that
the distribution is concentrated at $r_{\sub}=0.5$, despite of
a broadening due to finite-size effect, which
is more significant for larger $p$.

This subgraph resolves the ambiguity about GHZ entanglement mentioned in Sec.~\ref{subsec:ghz_ambiguity}. It also
explains the homogeneous long-range AED observed in 
Sec.~\ref{subsec:homoAED}. According to the analysis of solely the
$C_2$ toggles, edges from outside contribute only $O(N^{1/2})$ per 
vertex, yielding $O(N^{3/2})$ total, which normalizes to $O(N^{-1/2})$ and 
hence vanishes in the large-$N$ limit.
In other words, this dense subgraph is the hallmark of
the volume-law phase and $N_{\sub}/N$ can be viewed as an order of parameter
for MIPT. It may have two more implications:
\begin{enumerate}
\item Perhaps not many people have questioned, but it strongly
suggests that MIPT is a phase transition rather than a crossover,
considering the dependence of steady-state long-range AED $r$ on $p$ will 
experience a leap. The same reasoning also applies to dynamics.
A dynamical phase transition is observed for 
the emergence of GHZ entanglement~\cite{Xu2025prb}: AED is build up
from zero gradually but once it crosses the critical scaling $N^{-1/2}$
the subgraph $G(N_{\mathrm{sub}}, 1/2)$ suddenly appears and brings 
GHZ entanglement. In the Supplemental Material~\cite{sp} we recoded 
two movies of the time evolution of the adjacency matrix, one for
$\langle\Gamma_{i,j}\rangle$ and the other for an instance of
state trajectory, with $p=0.08$, $N=200$, and depth 600. From
the movies we can clearly see the formation and growth of a 
dense sub-matrix.
\item 
The subgraph is $G(N_\sub, 1/2)$ rather than any other
$G(N'_\sub, r<1/2)$. It implies that 
the output state of a monitored random Clifford circuits can be
understood as the output of a unitary Clifford circuits over $N_\sub$ qubits
(see Sec.~\ref{sec:key_observation}),
which are then 
perturbed by $N-N_\sub$ qubits with little entanglement. 
This result directly relates the output of monitored circuits to
unitary circuits, making the former much more transparent than
all previous treatment. As an important example,
it has been proved
in Ref.~\cite{Brown:2013aa} that, with $O(1)$ probability,
short all-to-all
random Clifford circuits (equivalent to deep
brickwork circuits studied here) produce 
\emph{good quantum error
correction code}, whose code space and
code distance are both proportional to the qubit number $N_{\sub}$.
This must be the physics underlying the series of discoveries about the
error-correction properties of the 
volume-law phase~\cite{Li:2021aa,Gullans:2020aa}. One may use this insight
to gain more thorough understandings.
\end{enumerate}

More remarks on the second point. In Ref.~\cite{Li:2021aa}, a concept of 
\emph{contiguous code distance} is employed to replace the
standard code distance. This seems to be the best we can do
in the framework of clipped gauge and KPZ equation. Our result indicates that
the error-correcting-capability is actually much stronger than considered previously,
due to the embedded good quantum error correction code,
especially if it is possible to conveniently tell which qubit is in the dense
subgraph according to the realizations of the circuit.
On the other hand, it does remind us to consider 
the spatial arrangement of the vertices of the
dense subgraph along the 1D chain.

\subsubsection{Clustering of Dense Vertices}

For convenience, we shall refer to the vertices forming this dense subgraph 
as dense vertices, and the rest as sparse vertices.  
We have seen an example of this arrangement in
Fig.~\ref{fig:graph_matrix}(a) where dense vertices are colored.
A key quantity of interest is the 
typical cluster size of dense vertices, where a cluster is defined as a 
contiguous block of dense vertices. 

To provide a reference point, we first examine a fully random 
distribution, in which every vertex is independently designated as dense with 
probability $\rho_{\sub}= N_{\sub}/N$. 
Fixing the left end of a dense cluster,
the probability for this cluster to have length $\ell$ is
$\rho_{\sub}^\ell (1-\rho_{\sub})^2$, i.e., a contiguous block of $\ell$ dense
vertices enclosed by two sparse vertices. Due to translation 
invariance, in the large-$N$ limit (hence no boundary effects)
the expected number of length-$\ell$ cluster is therefore
$\mathbb{E}[n_\ell] \sim N \rho_{\sub}^\ell (1-\rho_{\sub})^2$. We normalize it to
$p(\ell)= \rho_{\sub}^{\ell-1}(1-\rho_{\sub})$.

\begin{figure}[b]
    \centering
    \includegraphics[width=0.95\linewidth]{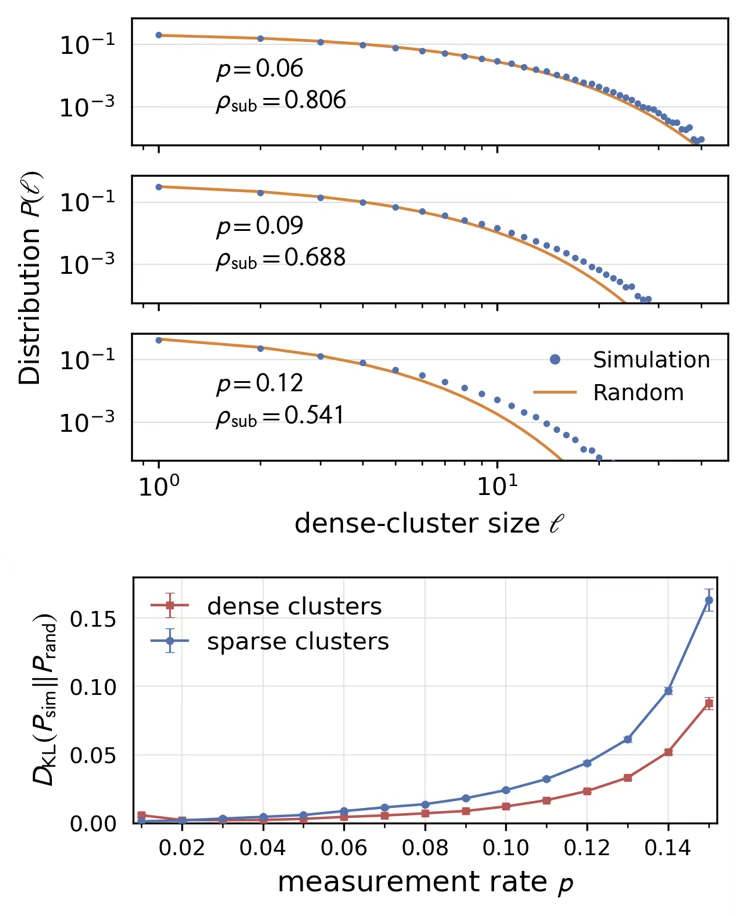}
    \caption{(a) Cluster-size distribution of the dense vertices 
    compared with random distribution for three values of $p$. (b)
    The deviation between them, for both the dense and the sparse vertices, 
    measured by the Kullback--Leibler 
    divergence.}
    \label{fig:vertex_distribution}
\end{figure}

In Fig.~\ref{fig:vertex_distribution}(a) we plot this distribution
and compared it with monitored random Clifford circuits for 
three different values of $p$. 
The probability of observing larger clusters is higher than that expected from 
a purely random distribution, indicating a clustering tendency among dense 
vertices. This clustering becomes more pronounced as 
$p$ increases. To demonstrate the discrepancy w.r.t.
random distributions, we calculate the relative entropy, or
Kullback--Leibler divergence between the two distributions (for both 
dense and sparse vertices) in Fig.~\ref{fig:vertex_distribution}(b).
It clearly shows curves rising with $p$.
This fact can be understood as follows: A larger $p$
implies a greater number of sparse vertices, thereby providing more 
configurational freedom for dense vertices to either 
group together or stay apart. 
The next question is thus how to understand this clustering. 
Any deviation from the random distribution signals the presence of an 
effective interaction between dense vertices. 
To capture this effect, we 
propose a toy model for monitored random circuits 
that incorporates a minimal dense–sparse ``interaction'', 
which is sufficient to reproduce the observed clustering behavior.

\subsection{Infection-Recovery Toy Model}
\label{subsec:toymodel}

We aim to devise a toy model that captures the 
clustering behavior and even shed light on MIPT 
from the side of volume-law phase. 
Previously we have confirmed that
there are two species of vertices. Therefore, 
a bold yet appealing idea is to model each qubit by a
classical bit in state labeled by either ``dense'' or ``sparse'', and 
transform the gates and measurements into flipping rules of 
the bits. This may sounds questionable but we find that it works to
capture some essential behaviors. 

\begin{figure}[tb]
    \centering
    \includegraphics[width=\linewidth]{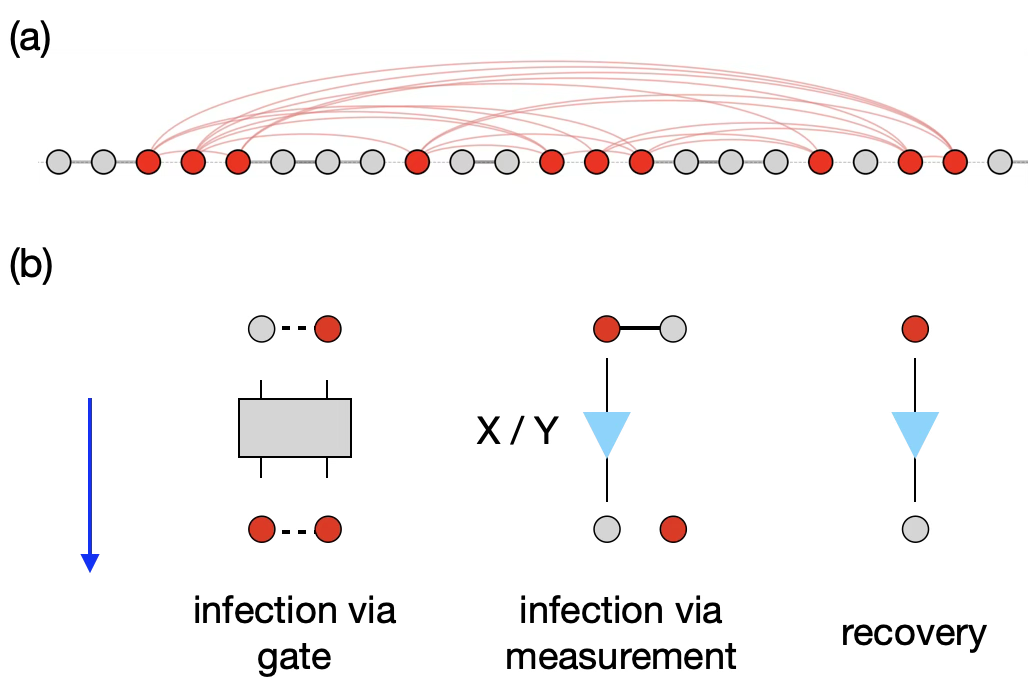}
    \caption{(a) Our toy model has two kinds of sites: red for 
    the $G(N_\sub, 1/2)$ subgraph and grey for sparse vertices. 
    (b) Schematic of the infection and recovery rules.}
    \label{fig:toy_rule}
\end{figure}

As shown in Fig.~\ref{fig:toy_rule}(a),
we use red and grey sites to represent the dense and sparse vertices, 
respectively. The red sites are mutually connected with 
probability 1/2 by definition.  
We model the grey sites as either isolated or
connected to one of its nearest neighbor 
in the 1D chain. Based on this setting, the edge toggles
summarized in Tab.~\ref{tab:1} give rise to 
three kinds of evolutions illustrated in Fig.~\ref{fig:toy_rule}(b).
We name our toy model by \emph{infection-recovery model}
because these evolutions are reminiscent of epidemic spreading in 1D space:
\begin{description}
    \item[Infection via gates] The effect of a random two-qubit Clifford gate 
    applied on a red--grey pair is to flip the grey site into red
    with probability $\alpha_u(p)$.
    \item[Infection via measurements] The local complementations employed in $X$- and $Y$-measurements on a 
    dense vertex (red site) will
    transform its sparse neighbor connected to it into red, while itself 
    becomes grey.
    \item[Recovery] $Z$-measurement will cut all edges from a dense vertex,
    making a red site grey.
\end{description}
Below we determine $\alpha_u(p)$ and explain why other configurations, such as
gates on red--red pairs, have no dynamical effect.

\subsubsection{Rate of Infection via Gates}

We consider the left configuration of Fig.~\ref{fig:toy_rule}(b)
and call the grey and red sites $v_{1}$ and $v_{2}$, respectively.
Referring to Tab.~\ref{tab:1}, when
$v_{1}$ and $v_{2}$ are connected $(v_1\sim v_2)$, we examine 
if $G^{P,Q}$ brings $v_{1}$ to dense:
\begin{description}
    \item[$(Z,Z)$] No, because adding or removing a single edge does not make a sparse vertex dense.
    \item[$(Z,X)$] Yes, because $\prod_{i\in\mu(2)}T_{1,i}$ connects $v_1$ to all the neighbors of $v_2$, lifting it into 
    a dense vertex. 
    \item[$(X,Z)$] No. 
    \item[$(Z,Y)$] Yes, similar to the case of $(Z,X)$.
    \item[$(Y,Z)$] No.
    \item[$(X,X)$] Yes but with a price. Since $v_1\in\mu^*(1)$ we have toggles $\prod_{j\in \mu(2)}T_{1,j}$ making
    $v_{1}$ dense. Meanwhile, however, we have $v_2\in\mu^*(1)$ hence
    toggles $\prod_{k\in \mu(2)}T_{2,k}$, which cut the edges incident to 
    $v_{2}$ converting it sparse. So the number of dense vertex does not change.
    \item[$(Y,X)$] Yes.
    \item[$(X,Y)$] Yes but with a price. Similar to the case of $(X,X)$,
    since $\mu^*(2)\triangle \mu^*(1)=\mu(2)$ there are toggles
    making $v_{2}$ sparse.
    \item[$(Y,Y)$] Yes but with a price. In
    $\prod_{(i,j)\in\mu^*(2)}T_{i,j}$ there are toggles
    $\prod_{j\in\tilde{\mu}(2)} T_{1,j}$ with $\tilde{\mu}(2)\equiv\mu(2)\setminus \{v_1\}$, which make $v_1$ dense.
    Meanwhile, the toggles $\prod_{j\in\tilde{\mu}(2)} T_{2,j}$
    reduces $v_{2}$ into a sparse vertex.
\end{description}
From the above we see only three of them induce net infection. 
Thus, we fix the gate-induced-infection rate at $\alpha_{\sim}=0.3$ conditioned on
$(v_1\sim v_2)$.

In the case of $(v_1\nsim v_2)$, $(Z,X)$, $(Z,Y)$ and $(Y,X)$
contribute net infection similarly. But the case of $(Y,Y)$ needs further
clarification, because the graph operation shown in Tab.~\ref{tab:1} 
is not symmetric w.r.t. the two qubits while the gate itself is. 
Using the numbering of Tab.~\ref{tab:1}, 
$v_1$ will be brought to dense whereas the opposite numbering 
will not. This does not affect the formalism of
canonical extended graph state, as we 
discussed in the end of Sec.~\ref{subsec:table}. However,
it yields an ambiguity for our toy model: The
gate-induced-infection rate $\alpha_{\nsim}$ 
in the case of
$(v_1\nsim v_2)$ can be either 0.3 or 0.4. In hindsight, 
if we choose a middle point 0.35, the toy model 
yields a density $\rho_{\sub}=N_{\sub}/N$ close to 
the simulation results, see Fig.~\ref{fig:toy_model}(a).
But here $p$ cannot be too close to the critical point, as we find the
absorbing-state phase transition may occur with $p\approx 0.15$.
On the other hand, choosing 0.4 leads to worse
agreement about $\rho_{\sub}$, but gives us a 
critical point between 
0.161 and 0.162, see Fig.~\ref{fig:toy_model}(b).

At last, we need to determine the connectivity of the nearest-neighboring pairs.
We do not incorporate the dynamics of this connectivity into our toy model,
in order to make the model as simple as possible. Instead, we manually fix this probability by  Eq.~\eqref{eq:r1}. 
Thus, the rate of infection by gates is
$\alpha_u(p)=\alpha_{\sim}\langle\Gamma_1\rangle+\alpha_{\nsim}(1-\langle \Gamma_1\rangle)$, where
$\langle\Gamma_1\rangle$ depends on $p$ implicitly.

\subsubsection{Summary of the Toy Model}

To summarize, we consider a 1D chain of $N$ sites.
The dynamics is controlled by a single parameter $p\in[0,1]$, which tunes the recovery probability.
We sweep over the chain simulating the brickwork circuits.
If a  nearest-neighboring pair have opposite colors, 
with probability $\alpha_u(p)$ the grey 
site is colored by red. Then, each 
site is examined individually. If $v_i$ is red, it recovers (becomes grey)
with probability $p$. 
Conditioned on this recovery, to incorporate infection via measurements,
we check if the right nearest-neighbor $v_{i+1}$ is grey. If
it is, we turn it to red with a probability of
$\frac{2}{3}\,\langle\Gamma_1(p)\rangle$; Otherwise we
implement this to the left nearest-neighbor $v_{i-1}$.

Here are some remarks.
When a two-qubit gate acts on a pair of dense vertices, 
nothing happens in this toy model. This
is because all the nonlocal toggles are
fundamentally the local complementations.
Suppose the edge density among the vertices in
$\mu(v_1)$ is $r$. Then, local complementation 
at $v_1$ updates $r$ to $1-r$, hence does not change
edge density if it is $1/2$. 
This is exactly the situation we have.

Another remark, in our toy model
the process that a sparse initial graph becomes dense gradually
is not included, since we cannot faithfully incorporate 
this effect without sacrificing simplicity. 
As a consequence, the toy model effectively possesses an 
absorbing state where all vertices are sparse. However, 
there might be space of
parameters allowing a stable \emph{active phase}
with a non-zero density of dense vertices, so that the system will
exhibit an \emph{absorbing-state phase transition} 
in the thermodynamic limit.
Correspondingly, in finite-size systems, the active phase manifests 
itself as a long-lived 
quasi-stationary state. By initially seeding the system with 
a finite fraction of dense vertices, the dynamics 
can reach the quasi-stationary regime and remains there 
for a sufficiently long time.

\subsubsection{Numerical Results}

\begin{figure}[bt]
    \centering
    \includegraphics[width=\linewidth]{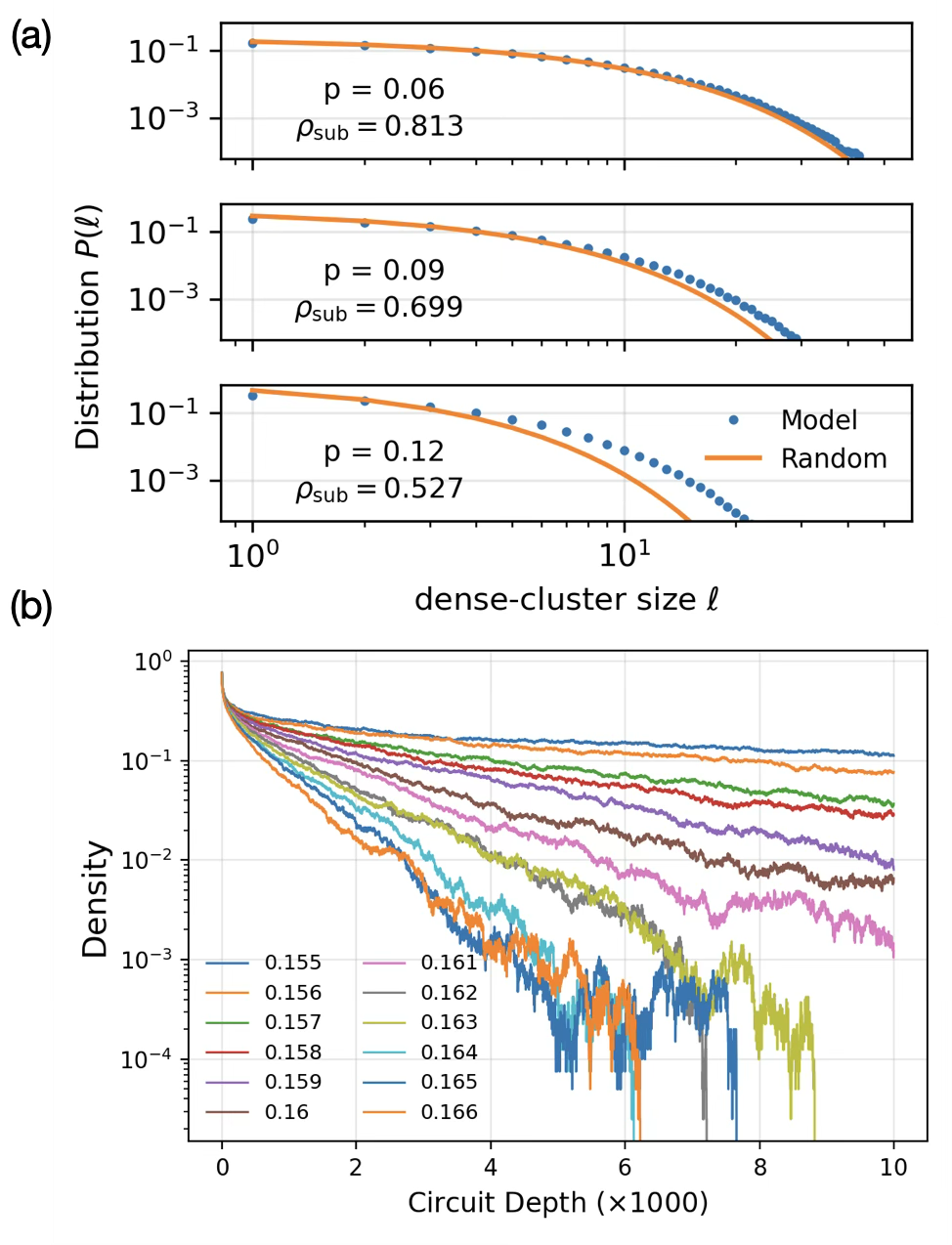}
    \caption{The infection-recovery model. (a) The distribution of
    the dense-vertex cluster sizes. $\rho_{\sub}$ denotes the density of red sites 
    in the quasi-stationary states, which agrees with the size of
    dense subgraph ($N_\sub/N$) given in Fig.~\ref{fig:graph_matrix}(b). 
    The results are obtained average over
    200 qubits, $10^{4}$ circuit levels, and 100 trajectories. 
    We have verified the (quasi) steady-state is well reached.
    (b) Density of red sites (dense vertices) for a system of 
    $N=200$ sites and different values of $p$ (averaged over 200 
    samples for each $p$). In this simulation, the rate of
    infection via gates is set at $\alpha_u(p)=0.4-0.1\langle\Gamma_1\rangle$.
    It shows an abrupt change of (quasi)-stationary state between $p=0.161$
    and $0.162$, suggesting an absorbing-state phase transition.}
    \label{fig:toy_model}
\end{figure}

We simulated the infection-recovery toy model with an initial
state with 100\% red sites and show the results of
(normalized) cluster-size distribution for three values 
of $p$ in Fig.~\ref{fig:toy_model}(a). More details
about the simulation is listed in the caption thereof. 
Clustering of red sites are demonstrated by a comparison
with random distributions. Comparing it Fig.~\ref{fig:vertex_distribution}(a),
we conclude that our infection-recovery model 
captures the essential part of 
the ``interaction'' between vertices inside and
outside of the Erd\H{o}s--R\'{e}nyi random subgraph.

Finally, we turn to the absorbing-state phase transition in this toy model. 
As noted earlier, setting the gate-induced infection probability for a 
pair of vertices to $0.4$ places the resulting transition remarkably close 
to the critical point $p_c \approx 0.16$ of the random Clifford circuits. 
A precise characterization of the critical behavior, however, requires 
dedicated work and lies beyond the scope of this paper. Here, we merely 
present the density of red sites (dense qubits) 
as a function of $p$ in Fig.~\ref{fig:toy_model}(b). 
The curve undergoes 
a sharp change between $p=0.161$ and $p=0.162$: The density drops abruptly 
to a vanishingly small value (for $N=200$, a value of order $\leq 10^{-4}$ 
corresponds to an exceedingly small probability), which strongly signals 
the occurrence of a phase transition. This observation suggests that 
MIPT can be interpreted through the lens of an absorbing-state phase
transition, offering a new perspective on the problem.

\subsection{$C_1$ Toggles and MIPT Critical Point}
\label{subsec:pc}

At last we discuss $C_1$ toggles and use it to analyze 
MIPT from the side of the area-law phase. In the area-law phase, the 
long-range AED is considerably small ($r\ll N^{-1}$) so that the effect of the 
$C_2$ toggles is characterized by $Nr^2\ll r$
according to Eq.~\eqref{eq:C2}. This makes them negligible compared with
the $C_1$ toggles at the scale of $r$.
Interestingly, while $C_1$ toggles are nonlocal w.r.t. the 1D qubit chain, 
they are \emph{local} w.r.t. the adjacency matrix entries as we explain below.

\begin{figure}[tb]
    \centering
    \includegraphics[width=0.7\linewidth]{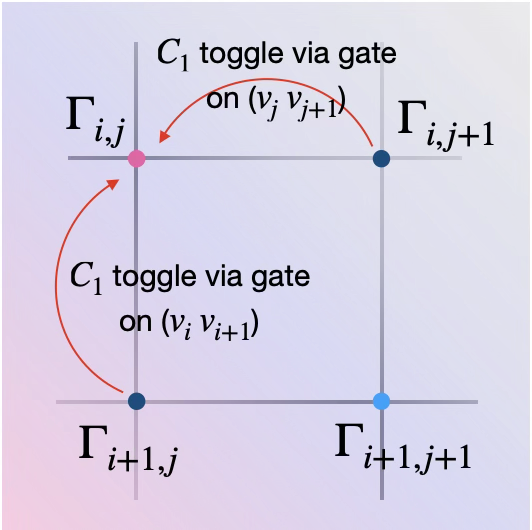}
    \caption{$C_1$ toggle as a local interaction between the entries of the 
    adjacency matrix. 
    The update of $\Gamma_{i,j}$ depends only on $\Gamma_{i,j+1}$ and 
    $\Gamma_{i+1,j}$, if the random gates 
    are implemented on the pairs $(v_j, v_{j+1})$
    and $(v_i, v_{i+1})$, respectively. The four entries shown in the figure comprise
    a closed system under $C_1$ toggles.}
    \label{fig:C1}
\end{figure}

Consider two vertices $v_i$ and $v_j$.
Without loss of generality, we assume the relevant gates of the current circuit
level are implemented on $(v_i,v_{i+1})$ and $(v_j,v_{j+1})$. 
Recall the introduction in Sec.~\ref{subsec:toggles},
a $C_1$ toggle realizing $\Gamma_{i,j}=0\rightarrow 1$  is 
conditioned on $\Gamma_{i,j+1}=1$ or $\Gamma_{i+1,j}=1$,
see the illustration in Fig.~\ref{fig:C1}.
The element $\Gamma_{i+1,j+1}$ automatically joints so that
the four elements comprise a closed system. One can verify this closeness 
by considering how the other three matrix entries update through
$C_1$ toggles (here we focus on $\Gamma_{i,j}$).
In the next circuit level, the gates are shifted to 
$(v_i,v_{i-1})$ and $(v_j,v_{j-1})$. Consequently,
$\Gamma_{i,j}$, $\Gamma_{i-1,j}$,
$\Gamma_{i,j-1}$ and $\Gamma_{i-1,j-1}$ will form a closed system instead.

To give a quick estimate of the critical point, we do
not aim at solving the complete distribution of edge probability but
exploit the fact that the long-range part must be bounded from above by 
some value $r\ll 1$.
Assuming $\langle \Gamma_{i,j}\rangle$ is $r$, let us calculate 
its value $\langle \Gamma'_{i,j}\rangle$ after one level of circuit.
Since $r\ll 1$, the probability of $\Gamma_{i,j}=1$ is the sum of all
different sources of possibilities. 
Thus, $\langle \Gamma'_{i,j}\rangle$ is $r$ plus the new chance
of $\Gamma_{i,j}=1$ enabled
by the $C_1$ toggles. Recall the toggle is conditioned on $\Gamma_{i,j+1}=1$
or $\Gamma_{i+1,j}=1$. To maximize  
$\langle \Gamma'_{i,j}\rangle$, we require $\langle \Gamma_{i,j+1}\rangle$ and 
$\langle \Gamma_{i+1,j}\rangle $ also have the saturated scale $r$. 
As a remark, the saturation scale of $\langle \Gamma_{ij}\rangle$
should decay with $d=\abs{i-j}$ in bulk of the area-law phase. But at 
the critical point it is valid to assume a homogeneous $r$.

In Sec.~\ref{subsec:toggles} we listed seven gates that may
contribute to the $C_1$ toggles. Now we need to identify
them more carefully. The $C_1$ toggle conditioned on $\Gamma_{i, j+1}=1$
must come from the gate applied on the ordered pair $(v_j, v_{j+1})$.
Using the notation of Tab.~\ref{tab:1}, we find
two gates when $(1\sim 2)$ [$(Z,X)$ and $(Z, Y)$] and four gates when $(1\nsim 2)$
[$(Z,X)$, $(Z,Y)$, $(Y,X)$, and $(Y,Y)$]
that have net positive contribution to the number of edges. 
Similar to the discussions in the infection-recovery model, some gates,
e.g., $(X,X):(1\sim 2)$, contributes
a $C_1$ toggle on $\Gamma_{ij}$ with the price of
other existing edges. Therefore, we do 
not consider such gates in the mean-field argument.
As in the discussion of the infection-recovery toy model, a 
special clarification is necessary for the 
case of $(Y,Y)$ gate. But
now the degrees of the two vertices $(v_1, v_2)$ 
have the same scale (the classification of dense and sparse does not apply here). 
Thus, we randomly determine the ordering, leading to a probability
0.35 in the case of $(1\nsim 2)$. 

After these preparations, we are ready to write down the
equilibrium condition for $\langle\Gamma_{i,j}\rangle$.
In order to have $\langle \Gamma_{i,j+1}\rangle=\langle \Gamma_{i+1,j}\rangle=r$,
both $v_{i+1}$ and $v_{j+1}$ must have not been measured in the
previous circuit level. The probability for this to happen is
$(1-p)^2$. Conditioned on this, the probability that
$(v_i\sim v_{i+1})$ and $(v_j\sim v_{j+1})$ is no longer
$\langle\Gamma_1\rangle$ given by Eq.~\eqref{eq:r1}. 
Instead it should be the conditioned
version $\gamma_1=\langle\Gamma_1\rangle/(1-p)^2$, because 
isolated vertex surely 
do not contribute to $\langle\Gamma_1\rangle$.
Therefore, the probability that a gate
makes $\Gamma_{i,j}=1$ via a $C_1$ toggle is
$0.2\gamma_1+0.35(1-\gamma_1)$.
Therefore, $\langle\Gamma'_{i,j}\rangle$ is estimated by
$r+2r[0.2\gamma_1+0.35(1-\gamma_1)](1-p)^2$.
At last,
random measurements keep only $(1-p)$ part of vertices,
hence $(1-p)^2$ edges, as the case of $C_0$ toggles discussed in Sec.~\ref{subsec:C0}.
Assembling all these together, we obtain the equilibrium condition 
\begin{equation}
    \left[1+\frac{7-3\gamma_1}{10}(1-p)^2\right](1-p)^2=1,
\end{equation}
which has a unique solution of $p= 0.1608$ in the interval 
$0\leq p\leq 1$. This is surprisingly consistent with the
numerical result of $p_c\approx 0.16$.

\section{Conclusions and Discussions}
\label{sec:conclusion}

To conclude, we introduced a graph-theoretic framework for 
monitored random Clifford circuits. 
Our approach departs from existing methods fundamentally: Instead of computing bipartite entanglement measures, 
we directly characterize the output states through their graph representations. 
This perspective enabled us to address questions that have remained inaccessible to the 
replica trick and the clipped gauge formalism.
Our main results are listed as following:
\begin{enumerate}
\item 
We established that uniformly random stabilizer states 
correspond asymptotically to Erd\H{o}s--R\'{e}nyi random graphs $G(N,1/2)$, which allowed 
us to derive the statistics of GHZ entanglement and resolve the even--odd effect analytically. 
\item
We demonstrated that the graph-update rules reduce to three 
types of edge toggles. This observation
helps us to quickly capture a series of important features. 
\item
We found that the nonlocal $C_2$ toggles lead to an emergent subgraph
$G(N_{\sub},1/2)$ with $N_{\sub}/N\approx \sqrt{1-p/p_c}$ for
a typical output state in the volume-law phase. 
Qubits out of this subgraph are sparsely entangled.
Thus, most distinguished
properties of the volume-law phase, including the quantum error correction
capability and the plateaus of GHZ entanglement, can be directly understood from
unitary circuits on $N_{\sub}$ qubits.
\item 
In the volume-law phase, there are two species of qubits, either
``dense'' or ``sparse.''  This leads to a toy model where the circuit
dynamics is modeled by classical infection and recovery.
The model is perhaps over-simplified but captures the clustering
of dense vertices and incorporates MIPT in the form of an absorbing-state phase transition.
\item 
We show that $C_1$ toggles are actually local interactions between
the entries of graph adjacency matrix. Then, a mean-field treatment of 
the $C_1$ toggles leads to an estimate $p_c = 0.1608$, in excellent agreement with numerical simulations.
\end{enumerate}

Our results demonstrate that the graph-based approach 
offers a versatile and powerful analytical toolkit, capable of 
addressing problems that have so far resisted theoretical understanding.
Several directions merit further investigation. 
One may consider graph-theoretical
conceptions, such as the maximum connected subgraph and the transition of
graph rank (in the field of real number)~\cite{Costello:2008aa}, 
and examine if they have physical meaning
in dynamics build upon Clifford circuits. 
Another example is that in the deep area-law phase, the whole graph
may split into disconnected components, the phenomenon
that captures the multipartite entanglement probed by quantum fisher information~\cite{Lira-Solanilla:2025aa}.
One may also generalize our formalism to include
magic. Moreover, while this Article mainly focus on the volume-law
phase, it is interesting and also important to study the typical output states
at the critical point of MIPT, where a Clifford conformal field theory has 
been suggested~\cite{Li2021Conformal}. It has been found that
the measurement-induced criticality supports multipartite GHZ entanglement,
which is statistically invisible 
in the volume-law phase~\cite{Xu2025prb}.

\begin{acknowledgements}
Y.-X. Z. acknowledges the financial support from the National Natural
Science Foundation of China (Grant No.~12375024), and the
CAS Project for Young Scientists in Basic Research (Grant
No.~YSBR-100).
\end{acknowledgements}

\appendix
\setcounter{figure}{0}
\setcounter{table}{0}
\renewcommand{\thefigure}{A\arabic{figure}} 
\renewcommand{\thetable}{A\arabic{table}}

\section{GHZ entanglement in graph states of Erd\H{o}s--R\'{e}nyi Random Graph $G(N, r\neq 1/2)$}
\label{app:Gnp}

As mentioned in the main text, the homogeneous long-range AED
cannot distinguish Erd\H{o}s--R\'{e}nyi random graphs $G(N'_{\sub}, r)$
with different $r$ but fixed $rN'_{\sub}/N$.
Here we show that GHZ entanglement statistics also fails to
distinguish between these ensembles.
Following the same trick used for the case of 
$r=1/2$ (unitary circuits), we start from examining the rank of 
the sub-matrix $\Gamma_A$.

\subsection{$\langle g_3\rangle=\langle \dim\ker\,\Gamma\rangle$ is still valid}

Here we show that the probability for $\Gamma_A$ to be full rank 
converges to 1 in the large-$N$ limit,
provided that the triangle inequality is satisfied. We define the
thermodynamical limit by diverging $N$ while holding a constant ratio 
$N_A/(N_B+N_C)=\alpha<1$ 

As same as in the case of $r=1/2$, we just need to show 
the $N_A\times (N_B+N_C)$ sub-matrix $\Gamma'_A$ is full rank
with 100\% probability. Its rows are linearly independent 
if and only if there exists a non-zero vector $\vec{v}\in \mathbb{F}_2^{N_A}$
so that $\vec{v}\cdot\Gamma'_A=0$. Then, by Markov's inequality,
the probability for it to \emph{not} have 
the full row rank is upper bounded by the expected number of such $\vec{v}$,
\begin{equation}
\label{eq:bound-r}
\begin{split}
    \mathrm{Pr}[\rank(\Gamma'_A)<N_A] & \leq \left\langle 
    \#\{\vec{v}\neq 0 \vert \vec{v}\cdot\Gamma'_A=0\}\right\rangle  \\
    & = \sum_{w=1}^{N_A}\binom{N_A}{w} p_w^{N_B+N_C}.
\end{split}
\end{equation}
In the second equality, $w$ denotes the Hamming weight of a 
vector $\vec{v}\in \mathbb{F}_2^{N_A}$, $\binom{N_A}{w}$ counts the 
number of such vectors, and $p_w^{N_B+N_C}$ denotes the probability 
that all columns of $\Gamma'_A$ are orthogonal to $\vec{v}$. Hence,
the $p_w$ is the probability to have even number of $1$
at the coordinates where $\vec{v}$ is 1:
\begin{equation}
    p_w=\frac{1+(1-2r)^{w}}{2}.
\end{equation}
The target is to show the right hand side of Eq.~\eqref{eq:bound-r} vanishes in 
the large-$N$ limit for any fixed $r$.

Next, we split the sum in \eqref{eq:bound-r} into two regimes:
Small weight ($w \le W$) and large weight ($w > W$),
with 
\begin{equation}
W = \lceil \frac{\log_2(N_B+N_C) }{-\log_2 (1-2r)} \rceil.
\end{equation}

\subsubsection{Small weight $1 \le w \le W$}
For such $w$ we have $W = O[\log_2 (N_B+N_C)]$ and $(1-2r)^w < 1$, hence
$p_w \le \frac{1+1-2r}{2} = 1-r$.
The binomial coefficient satisfies 
\begin{equation} 
\binom{N_A}{w} \le N_A^w \le (N_A)^W
= 2^{\,O[(\log_2 N_A)^2]}.
\end{equation}
It leads to
\begin{equation}
\begin{split}
\binom{N_A}{w} (p_w)^{N_A/\alpha} & \le\; (N_A)^W (1-r)^{N_A/\alpha} \\
& = 2^{\,O((\log N_A)^2) - c_1 N_A/\alpha},
\end{split}\end{equation}
where $ c_1 = -\log_2(1-r) > 0 $.
Therefore, we have
\begin{equation}
\begin{split}
\sum_{w=1}^{W} \binom{N_A}{w} (p_w)^{N_A/\alpha}
& \le\; W \cdot (N_A)^W 2^{-c_1 N_A/\alpha} \\
& = 2^{-c_1 N_A/\alpha + O((\log_2 N_A)^2)} \\
& \xrightarrow{\text{large}\, N}\; 0 .
\end{split}
\end{equation}

\subsubsection{Large weight $W < w \le N_A$}
By definition of $W$, $(1-2r)^w \le (1-2r)^W \le 1/(N_B+N_C)$.
Hence $\varepsilon_w := (1-2r)^w$ satisfies $0 \le \varepsilon_w \le 1/(N_B+N_C)$, and
\begin{equation}
p_w = \frac12(1+\varepsilon_w).
\end{equation}
Using $\ln(1+u) \le u$, we obtain
\begin{equation}
\begin{split}
-\log_2 p_w & = 1 - \log_2(1+\varepsilon_w)
            \ge 1 - \frac{\varepsilon_w}{\ln 2} \\
            & \ge 1 - \frac{1}{(N_B+N_C)\ln 2}.
\end{split}
\end{equation}
Consequently,
\begin{equation}
(p_w)^{N_B+N_C} \;\le\;
        2^{-(N_B+N_C) + 1/\ln 2}.
\end{equation}
For the binomial, we use the exponential entropy bound
\begin{equation}
\binom{N_A}{w} \le 2^{N_A H(w/N_A)}
\end{equation}
where $H(t) = -t\log_2 t - (1-t)\log_2(1-t)$
is the binary entropy.  
Then we have
\begin{align*}
\binom{N_A}{w} (p_w)^{N_B+N_C}
&\le 2^{N_A H(w/N_A) - N_A/\alpha + 1/\ln 2} \\
& = 2^{-N_A\,[1/\alpha-H(w/N_A)] + 1/\ln 2}.
\end{align*}
Recall $H(t) \le 1$ for all $t\in[0,1]$,
which means $1/\alpha-H(w/N_A)>1/\alpha-1>0$.
Hence, the above formula is suppressed exponentially.
Then, it follows that 
\begin{equation}
\begin{split}
    \sum_{w=W+1}^{N_A} \binom{N_A}{w} (p_w)^{N_B+N_C}
&\le\; N_A \cdot 2^{-N_A(1/\alpha-1) + 1/\ln 2} \\
& \xrightarrow{\text{large}\, N}\; 0 .
\end{split}
\end{equation}
Assembling the above two parts finishes the proof.

\subsection{Universality of the nullity distribution for
Erd\H{o}s--R\'enyi random graphs}

We learn that the nullity distribution is 
$r$-independent from a universality theorem of Nguyen and 
Wood for random skew-symmetric matrices~\cite{nguyenLocalGlobalUniversality2025}. Their 
result concerns integer-valued skew-symmetric matrices whose upper-triangular entries are 
independent and identically distributed copies of an integer-valued random variable $\xi$. 
The variable $\xi$ is said to be $\alpha$-balanced if there exists a constant $\alpha>0$, 
independent of the matrix dimension, such that, for every prime $\ell$ and every residue 
class $a\in\mathbb{Z}/\ell\mathbb{Z}$,
	\begin{equation*}
		\Pr\!\bigl(\xi\equiv a \pmod{\ell}\bigr)\leq 1-\alpha.
	\end{equation*}
The final part of Theorem~1.13 of Ref.~\cite{nguyenLocalGlobalUniversality2025} states that, under this condition, for any fixed prime $\ell$, the rank probabilities of the matrix reduced modulo $\ell$ converge to universal values in the large-$N$ limit.

To apply this result to graph adjacency matrices, define the integer-valued skew-symmetric matrix
\begin{equation*}
    (A_N)_{ij}=
    \begin{cases}
        (\Gamma_N)_{ij}, & i<j,\\
        -(\Gamma_N)_{ji}, & i>j,\\
        0, & i=j.
    \end{cases}
\end{equation*}
Since $-1=1$ in $\mathbb{F}_2$, reduction modulo $2$ gives
\begin{equation*}
    A_N \bmod 2 = \Gamma_N.
\end{equation*}
The independent upper-triangular entries of $A_N$ follow a Bernoulli distribution,
\begin{equation*}
    \Pr(\xi=1)=r,
    \qquad
    \Pr(\xi=0)=1-r.
\end{equation*}
For any prime $\ell$, the residue classes $0$ and $1$ occur with probabilities $1-r$ and $r$, respectively, while all other residue classes occur with probability zero. Hence, for $0<r\leq 1/2$,
\begin{equation*}
    \max_{a\in\mathbb{Z}/\ell\mathbb{Z}}
    \Pr\!\bigl(\xi\equiv a \pmod{\ell}\bigr)=1-r.
\end{equation*}
Thus, $\operatorname{Bernoulli}(r)$ is $r$-balanced, and the Nguyen--Wood theorem applies.

\begin{figure}[bt]
  \centering
  \includegraphics[width=\linewidth]{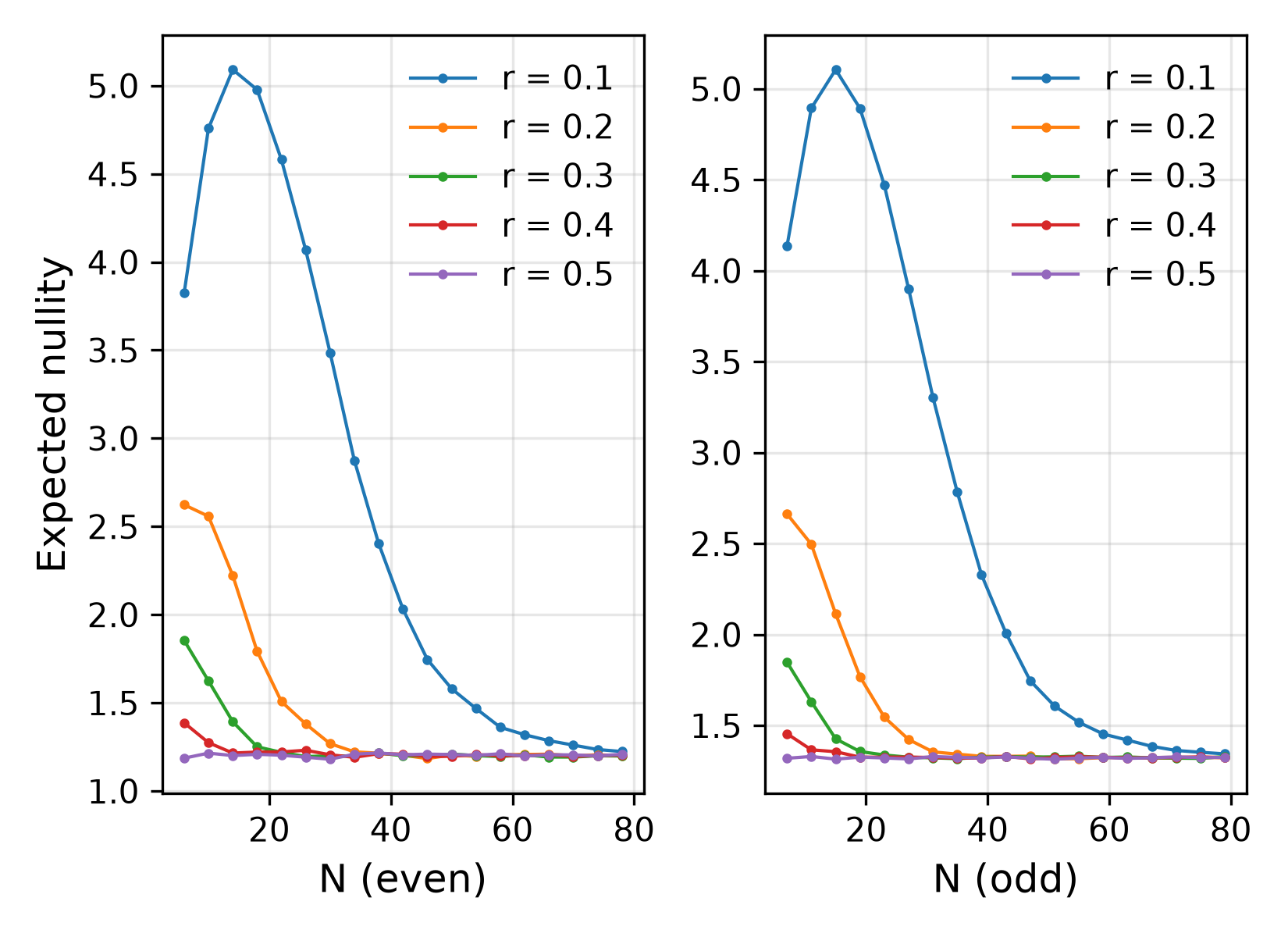}
  \caption{Expected nullity of Erd\H{o}s--R\'{e}nyi graphs $G(N,r)$ for 
  $r=0.1,0.2,\ldots,0.5$ and even/odd $N$ in the left/right panel. 
  The curves collapse in the large-$N$ limit.}
  \label{fig:Gnr}
\end{figure}

To verify this result, we sampled Erd\H{o}s--R\'{e}nyi graphs $G(N, r)$ with $r$ ranging from $0.1$ to 
$0.5$ in steps of $0.1$, and evaluated the expected nullity. The results, shown 
in Fig.~\ref{fig:Gnr}, indeed confirm that in the thermodynamic limit $N\to\infty$, 
the expected nullity approaches a universal constant, irrespective of $r$.

\section{Brief Review of the Graph Operations}
\label{app:measurements}

Here we outline the derivation of the graph operations 
corresponding to the gates $G^{P,Q}$ and the Pauli measurements,
and also the graph operations to recast an extended graph state into
the canonical form introduced in Ref.~\cite{Hu:2022aa}.
For later convenience, we start from the measurements.

\subsection{$Z$--measurement}
This is the simplest case. Recall the stabilizer generators
of graph states~\eqref{eq:graph-stabilizer}. 
$Z$--measurement on qubit $v_1$ commutes with stabilizer generations
defined on all other vertices. Thus, if the measurement outcome is $+1$ 
the corresponding
graph operation is to isolate $v_1$ from all other vertices.
If the measurement outcome is $-1$, then
\begin{equation}
    \mathcal{S}_i \rightarrow - \mathcal{S}_i, \;\forall i\in\mu(1).
\end{equation}

\subsection{$Y$--measurement}

The standard stabilizer generators are spelled by $X$ and $Z$ 
operators. We can multiply  $\mathcal{S}_1$ given in Eq.~\eqref{eq:graph-stabilizer}
by $\mathcal{S}_i$ for every $v_i\in\mu(1)$ to obtain
stabilizers having $Y_1$ explicitly:
\begin{equation}
\tilde{S}_i=\mathcal{S}_{1}\mathcal{S}_i= Y_1 Y_i\bigotimes_{j\in \mu(1)\triangle \mu^*(i)}Z_j.
\end{equation}
These stabilizers commute with $Y_1$, hence survive from the measurement of $Y_1$,
while $\mathcal{S}_1$ is replaced by $\pm Y_1$.
To convert them into the standard graph-state stabilizers, additional local
Clifford gates are required to map $Y_i$ into $X_i$.
The above formula also tells us what the new graph looks like.
We need to first apply a \emph{local complementation} around $v_1$,  and
secondly isolate $v_1$.

\subsection{$X$--measurement}
Now $\mathcal{S}_1$ commutes with the measurement so that the
corresponding post-measurement
stabilizer reads
\begin{equation}
S_1\rightarrow \pm\bigotimes_{j=1}^{N} Z_j^{\Gamma_{1,j}},
\end{equation}
where the sign depends on the measurement outcome.
For any $v_i\in \mu(1)$ the corresponding stabilizer has a factor
of $Z_1$ which does not commute with the measurement, hence,
needs extra treatment.
We pick up an arbitrary vertex $v_2\in\mu(1)$ 
and multiply $\mathcal{S}_i$ by $\mathcal{S}_2$ to obtain 
\begin{equation}
S'_i= X_i X_2 (Z_i Z_2)^{\Gamma_{i,2}}
\bigotimes_{j}Z_j^{\Gamma_{i,j}+\Gamma_{2,j}}
\end{equation}
which is trivial on qubit $v_1$, hence survives from the measurement.

To convert it to the standard graph-state form, we apply 
a Hadamard on $v_2$, hence, 
\begin{equation}
S'_1 \rightarrow \pm X_2 \bigotimes_{j\neq 1,2} Z_j^{\Gamma_{2,j}}    
\end{equation}
and 
\begin{equation}
S'_i\rightarrow     
\begin{cases}
        X_i Z_2 \bigotimes_{j}Z_j^{\Gamma_{i,j}+\Gamma_{2,j}}, & \Gamma_{i,2}=0; \\[0.7em]
        Y_i Y_2 \bigotimes_{j\neq 2,i}Z_j^{\Gamma_{i,j}+\Gamma_{2,j}}, & \Gamma_{i,2}=1.
    \end{cases}
\end{equation}
Above, the case of $(v_i\nsim v_2)$ is already in the standard
graph-state form. For the case of  $(v_i\sim v_2)$, 
we multiply it by the new $S'_1$ and obtain
\begin{equation}
S'_i\rightarrow \pm X_iZ_2\bigotimes_{j\neq 2,i} Z_j^{\Gamma_{1,j}.
+\Gamma_{i,j}+\Gamma_{2,j}}.
\end{equation}
The new graph thus can be determined referring to these stabilizers.
The result is that these operations can be 
summarized into an \emph{edge complementation}
(with respect to an edge connecting $v_1$ and $v_2$): It is
    equivalent to three local complementations: ${v_1}\circ {v_2}\circ {v_1}(\Gamma)$, or equivalently switching $v_1$ and $v_2$.

\subsection{Graph operations for Gates}

The new state $G^{P,Q}\ket{\Gamma}$ can be expanded as 
\begin{equation}
\label{eq:gate-operation}
    \frac{1}{2}\left[(\mathbb{I}+P_1)+Q_2(\mathbb{I}-P_1)\ket{\Gamma}
    \right],
\end{equation}
where $(\mathbb{I}\pm P_1)$ is the projector onto 
the $\pm 1$ eigenstate of $P_1$. Stabilizers of the 
post-measurement states $(\mathbb{I}\pm P_1)\ket{\Gamma}$ 
are different by some signs. It means that the states themselves can
be connected by a sequence of $Z$ gates, using the fact that
$ZXZ=-X$. It turns out that Eq.~\eqref{eq:gate-operation} can 
always be
shaped into the form of
\begin{equation}
\begin{split}
    &\frac{1}{\sqrt2}\left(
    \mathbb{I}+i^k\prod_{j\in B} Z_j\right)\ket{G} \\
    = & 
    H_1 Z_1 \prod_{x,y \in \mu^*(1)}\mathrm{CS}^k_{x,y} \prod_{x\in \mu^*(1), y\in B}
    \mathrm{CZ}_{x,y}\ket{G}
    \end{split}
\end{equation}
where $\ket{G}$ is some grpah state, 
$B$ is a set including $v_1$, $\mathrm{CS}_{x,y}$ is the Controlled-S gate
$\ket{0}_x\bra{0}\otimes\mathbb{I}_y+\ket{1}\bra{1}\otimes S_y$, and
$\mathrm{CS}_{x,x}=S_{x}$ by convention.
The above equality is introduced in Ref.~\cite{Khesin:2021aa}.
It is the key knob for the graph operations listed in Tab.~\ref{tab:1}.

\subsection{Graph operations for the canonical form}

If the two ends of an edge are both assigned 
a Hadamard gate, or the vertex larger numbered is assigned a Hadamard,
extra graph operations are necessary to bring it into the canonical form.
This operation is based on 
the formula~\cite{Elliott:2008aa}
\begin{equation}
\label{eq:HH}
    H_a H_b\ket{\Gamma}= Z_a Z_b \prod_{i\in \mu^*(x)}\prod_{j\in \mu^*(y)} \mathrm{CZ}_{i,j}\ket{\Gamma},
\end{equation}
which is conditioned on $(v_a, v_b)$ is connected.

\bibliography{ReferencesGHZ.bib}

\end{document}